\documentclass[sigplan, screen]{acmart}

\usepackage{booktabs}   %% For formal tables:
\usepackage{subcaption} %% For complex figures with subfigures/subcaptions
\usepackage{graphicx}
\usepackage{pifont}
\usepackage{wrapfig}
\usepackage{hyperref}
\usepackage{fancyvrb}
\usepackage{xspace}
\usepackage{algorithmicx}
\usepackage{algpseudocode}
\usepackage{algorithm}
\usepackage{amsfonts}
\usepackage{amsmath}
\usepackage{amsthm}
\usepackage{amssymb}
\usepackage{semantic}
\usepackage{listings}
\usepackage{array}
\usepackage{fancyvrb}
\usepackage{mathpartir}
\usepackage{stmaryrd}
\usepackage{graphicx}
\usepackage{caption}
\usepackage{syntax, etoolbox}
\usepackage{tikz}
\usepackage{proof}
\usepackage{color}
\usepackage{mathtools}
\usepackage{adjustbox}

\theoremstyle{problem}

\theoremstyle{definition}

\lstdefinestyle{caddy}{
  basicstyle=\footnotesize\sffamily,
}
\lstdefinestyle{rewrite}{
  basicstyle=\scriptsize\sffamily,
  gobble=4,
}
\makeatletter
\lst@AddToHook{Init}{\mathligsoff}
\makeatother

\usepackage[utf8x]{inputenc}

\newcommand{\K}[1]{\ensuremath{\mathbf{#1}}\relax\ifmmode\:\else\fi}
\newcommand{\F}[1]{\text{\footnotesize{\textsf{#1}}}}

\newcommand{\shepherd}[1]{#1}

\newcommand{\tool}{Szalinski\xspace}

\newcommand{\rust}{Rust\xspace}
\newcommand{\cadilac}{Caddy\xspace}
\newcommand{\caddy}{Caddy\xspace}

\newcommand{\fold}{\F{Fold}\xspace}

\newcommand{\map}{\F{Map2}\xspace}

\newcommand{\maps}{\F{Map2}\F{s}\xspace}
\newcommand{\mapi}{\F{Tabulate}\xspace}
\newcommand{\mapis}{\F{Tabulate}\F{s}\xspace}
\newcommand{\rpt}{\F{Repeat}\xspace}
\newcommand{\rpts}{\F{Repeats}\xspace}

\newcommand{\unit}{\F{Cuboid}\xspace}

\newcommand{\cyl}{\F{Cylinder}\xspace}
\newcommand{\hx}{\F{Hexprism}\xspace}
\newcommand{\scale}{\F{Scale}\xspace}
\newcommand{\rotate}{\F{Rotate}\xspace}
\newcommand{\trans}{\F{Translate}\xspace}
\newcommand{\polar}{\F{TranslateSpherical}\xspace}
\newcommand{\union}{\F{Union}\xspace}
\newcommand{\unions}{\F{Unions}\xspace}

\newcommand{\slist}{\F{List}\xspace}
\newcommand{\slists}{\F{Lists}\xspace}

\newcommand{\concat}{\F{Concat}\xspace}

\newcommand{\unpolar}{\F{Unspherical}\xspace}
\newcommand{\unsort}{\F{Unsort}\xspace}
\newcommand{\ssort}{\F{Sort}\xspace}
\newcommand{\unpart}{\F{Unpart}\xspace}
\newcommand{\spart}{\F{Part}\xspace}
\newcommand{\egraph}{\mbox{E-graph}\xspace}
\newcommand{\eclass}{eclass\xspace}
\newcommand{\enode}{enode\xspace}
\newcommand{\egraphs}{\mbox{E-graphs}\xspace}
\newcommand{\eclasses}{eclasses\xspace}
\newcommand{\enodes}{enodes\xspace}
\newcommand{\Semtag}{Inverse transformations\xspace}
\newcommand{\semtag}{inverse transformations\xspace}
\newcommand{\semt}{inverse transformation\xspace}
\newcommand{\semts}{inverse transformations\xspace}

\newcommand{\subst}[3]{\ensuremath{#1[#3/#2]}}

\newcommand\rewritename[1]{
  \multicolumn{3}{l}{\footnotesize \textbf{\textsf{#1}}}
  \vspace{1mm}
}
\newcommand\rewritenamecond[2]{
  \multicolumn{3}{l}{\footnotesize \textbf{\textsf{#1}} \textsf{#2}}
  \vspace{1mm}
}
\newcommand\rewritespace{\vspace{2mm}}
\newcommand\rewritesto{\ensuremath{\rightsquigarrow}}
\newcommand\rewritesboth{\ensuremath{\leftrightsquigarrow}}

\newcommand\Rewritename[1]{
  \footnotesize \textbf{\textsf{#1}}
  \vspace{0mm}
}
\newcommand\Rewritenamecond[2]{
  \footnotesize \textbf{\textsf{#1}} \textsf{#2}
  \vspace{0mm}
}

\makeatletter
\newcommand\etc{etc\@ifnextchar.{}{.\@}}
\makeatother

\newcommand{\toolLinesRust}{3,000\xspace}
\newcommand{\toolNumRules}{65\xspace}
\newcommand{\thingBenchCount}{2,127\xspace}
\setcopyright{none}
\renewcommand\footnotetextcopyrightpermission[1]{}
\setcopyright{none}
\makeatletter
\renewcommand\@formatdoi[1]{\ignorespaces}
\makeatother

\begin{document}

%% Title information
\title[]{Synthesizing Structured CAD Models with \\ Equality Saturation and Inverse Transformations}

\author{Chandrakana Nandi}
\affiliation{
  \institution{University of Washington}            %% \institution is required
  \country{USA}                    %% \country is recommended
}
\email{cnandi@cs.washington.edu}          %% \email is recommended
\author{Max Willsey}
\affiliation{
  \institution{University of Washington}            %% \institution is required
  \country{USA}                    %% \country is recommended
}
\email{mwillsey@cs.washington.edu}          %% \email is recommended
\author{Adam Anderson}
\affiliation{
  \institution{University of Washington}            %% \institution is required
  \country{USA}                    %% \country is recommended
}
\email{adamand2@cs.washington.edu}          %% \email is recommended
\author{James R. Wilcox}
\affiliation{
  \institution{Certora}            %% \institution is required
  \country{USA}                    %% \country is recommended
}
\email{james@certora.com}          %% \email is recommended
\author{Eva Darulova}
\affiliation{
  \institution{MPI-SWS}            %% \institution is required
  \country{Germany}                    %% \country is recommended
}
\email{eva@mpi-sws.org}          %% \email is recommended
\author{Dan Grossman}
\affiliation{
  \institution{University of Washington}            %% \institution is required
  \country{USA}                    %% \country is recommended
}
\email{djg@cs.washington.edu}          %% \email is recommended
\author{Zachary Tatlock}
\affiliation{
  \institution{University of Washington}            %% \institution is required
  \country{USA}                    %% \country is recommended
}
\email{ztatlock@cs.washington.edu}          %% \email is recommended

\renewcommand{\shortauthors} {C. Nandi et al.}
% \renewcommand{\shortauthors} {C.Nandi, M. Willsey, A. Anderson, J. R. Wilcox, E. Darulova, D. Grossman, Z. Tatlock}

%% Abstract
%% Note: \begin{abstract}...\end{abstract} environment must come
%% before \maketitle command
\begin{abstract}
  Recent program synthesis techniques
  help users customize CAD models (e.g., for 3D printing)
  by decompiling low-level triangle meshes to
  Constructive Solid Geometry (CSG) expressions.
Without loops or functions,
  editing CSG can require
  many coordinated changes,
  and existing mesh decompilers use heuristics that
  can obfuscate high-level structure.

This paper proposes a second decompilation stage
  to robustly ``shrink'' unstructured CSG expressions
  into more editable programs with map and fold operators.
We present \tool,
  a tool that uses Equality Saturation % and \egraphs
  with semantics-preserving CAD rewrites
  to efficiently search for smaller equivalent programs.
\tool relies on \emph{\semtag},
  a novel way for solvers to speculatively add
  equivalences to an \egraph.
We qualitatively evaluate \tool in case studies,
  show how it composes with an
  existing mesh decompiler, and
  demonstrate that \tool can shrink
  large models in seconds.

\end{abstract}

%% 2012 ACM Computing Classification System (CSS) concepts
%% Generate at 'http://dl.acm.org/ccs/ccs.cfm'.
% <concept_id>10011007.10011006.10011008</concept_id>
% <concept_desc>Software and its engineering~General programming languages</concept_desc>
% <concept_id>10003456.10003457.10003521.10003525</concept_id>
% <concept_desc>Social and professional topics~History of programming languages</concept_desc>
% <concept_significance>300</concept_significance>

% \ccsdesc[500]{Software and its engineering~General programming languages}
% \ccsdesc[300]{Social and professional topics~History of programming languages}
%% End of generated code

%% Keywords
%% comma separated list
\keywords{Program Synthesis, Equality Graph, Decompilation, Computer-Aided Design}  %% \keywords are mandatory in final camera-ready submission

%% \maketitle
%% Note: \maketitle command must come after title commands, author
%% commands, abstract environment, Computing Classification System
%% environment and commands, and keywords command.
\maketitle

\AtBeginEnvironment{grammar}{\small}
%!TEX root = main.tex
\section{Introduction}
\label{sec:intro}

\newcommand{\chOne}{\textbf{(C1)}\xspace}
\newcommand{\chTwo}{\textbf{(C2)}\xspace}

The programming languages and machine learning communities
  have developed techniques to
  decompile Computer-Aided Design (CAD) models from
  low-level numerical representations to
  Constructive Solid Geometry (CSG) expressions~\cite{
    reincarnate, inverse, latex, shape, csgnet, sherman19, genetic}.
These techniques aim to help users
  modify designs shared in online repositories~\cite{
    chilana1, chilana2, thing}.

Recent program synthesis results~\cite{inverse, reincarnate}
  decompile \emph{meshes},
%  defined CAD decompilation as converting a low-level \emph{mesh},
  sets of triangles defining an object's surface,
  into equivalent CSG expressions.
CSG includes
  geometric primitives like cylinders,
  affine transformations like translate,
  and set theoretic operators like union.
%We refer to this early CAD decompilation challenge as ``Phase 1.''

%The output of existing mesh decompilers
%  (\autoref{fig:motiv}, left) is \emph{flat}:
%  CSG has no loops or functions.

\begin{figure*}[t]
  \includegraphics[width=\textwidth]{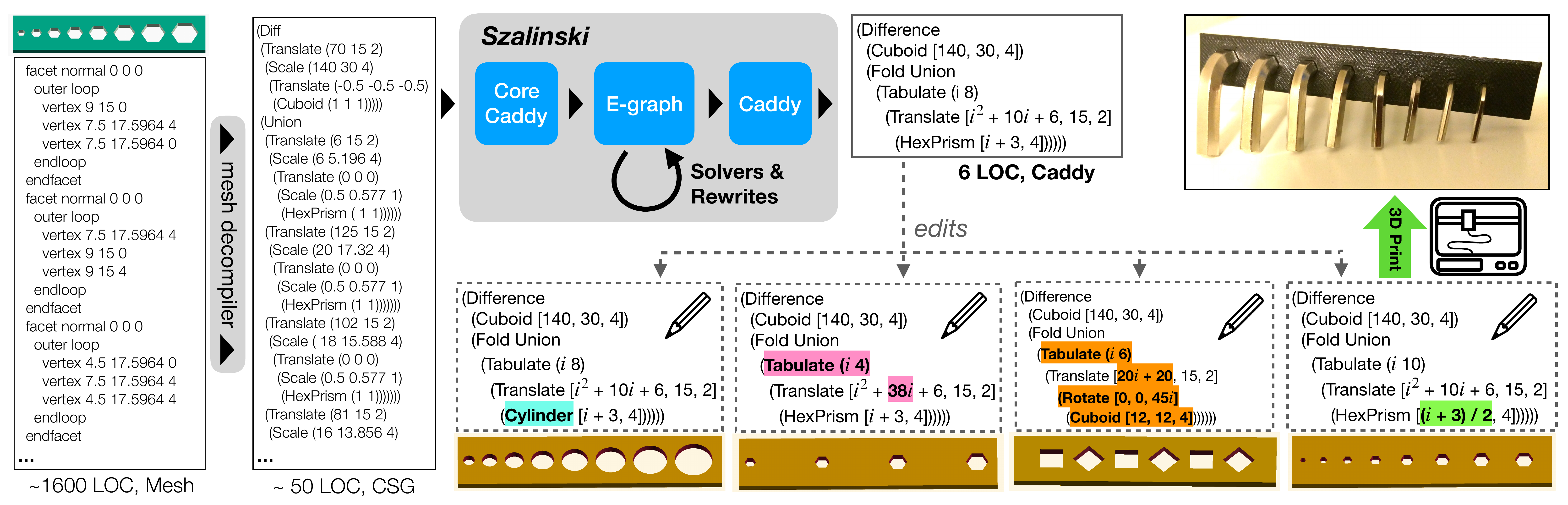}
  \caption{
    Existing mesh decompilers turn
      triangle meshes into CSG expressions.
%  The size of both mesh and CSG are roughly proportional to
%    the number of geometric features in the CAD model.
  \tool robustly synthesizes smaller, structured
    \cadilac programs from CSG expressions.
  This can ease customization by simplifying edits:
    small, mostly local changes
    yield usefully different models.
  The photo shows the 3D printed hex wrench holder after
    customizing hole sizes.
}
  \label{fig:motiv}
\end{figure*}

Existing mesh decompilers
  synthesize \emph{flat} output:
  CSG has no loops or functions (\autoref{fig:motiv}, left).
Therefore, CSG synthesized from
  large meshes with repetitive features
  also tends to be large and repetitive.
As in traditional programming,
  repetition makes otherwise intuitive edits
  tedious and error-prone.

Mesh decompilation is under-constrained~\cite{inverse, reincarnate}, so
  past tools rely on heuristics which cause them \shepherd{to
  exhibit two challenging features:
  \chOne synthesize equivalent but dissimilar CSG expressions for
  the same feature repeated under different transformations, and
  \chTwo arbitrarily order CSG subexpressions. These two
features, \chOne and \chTwo obfuscate
  high-level structure latent in synthesized CSG.}

This paper proposes a second decompilation stage
  that composes with prior work:
  given a flat CSG expression,
  produce an equivalent, smaller, and more editable program
  with map and fold operators for expressing repetition.
We present \emph{\tool}\footnote{The protagonist in the hit movie Honey I Shrunk the Kids was named Dr. Szalinski. Our work shrinks CADs rather than kids.}
(\autoref{fig:motiv}),
  a tool which combines semantics-preserving rewrites
  with simple solvers to
  synthesize structured CAD programs
  in a language called \emph{\caddy}.

\tool is designed to robustly handle the
  noisy and unstructured outputs of
  existing mesh decompilers.
In many of these outputs,
  high-level structure is only apparent after
  a set of CAD-specific rewrites
  have been judiciously applied \chOne.
Past work on Equality Saturation~\cite{eqsat}
  suggests that Equality Graphs (\egraphs)~\cite{nelson}---an
  efficient data structure underlying SMT solvers~\cite{z3, simplify}
  and program optimizers~\cite{denali, eqsat, eqsat1, carp}---would
  make a good fit for \tool
  because \egraphs can compactly encode many of the
  equivalent ways to express a program
  with respect to a set of rewrites.

Unfortunately,
  reordering with associative and commutative rewrites
  can cause \egraphs to blow up exponentially.
\shepherd{This is known as the
\textit{AC-matching problem}~\cite{acmatch1, acmatch2, acmatch3}.}
It presents a significant challenge for \tool because
  existing mesh decompilers typically output CSG features
  ordered by heuristics (e.g., geometric proximity)
  rather than high-level structure \chTwo.

To address the AC-matching problem in \tool
  we present \textit{\semtag},
  a novel way for solvers to
  \textit{speculatively} unify
  expressions in an \egraph
  which would be equivalent modulo
  reordering or partitioning.
Before unifying a result $R$ with its input $I$,
  a solver can annotate $R$ with an \semt which encodes
  how it manipulated $I$ to find
  the more-profitable $R$.
\tool then uses syntactic rewrites to
  \textit{propagate} and \textit{eliminate} \semts
  when opportunities to use such results arise.

To summarize, the contributions of this paper include:
\begin{itemize}
\item \emph{\tool}, a tool that takes
  a flat CSG expression as input and synthesizes
  a smaller equivalent program in
  \emph{\caddy}, a language that extends CSG with
  map and fold operators for expressing repetition.

\item \textit{\Semtag}, a new technique for
    interfacing simple-yet-effective
    structure finding solvers with \egraphs.
  The technique is not CAD-specific,
    but is particularly useful for
    reordering CAD operations.

% note: not citing shape and csgnet
\item A case study composing \tool with a
  recent mesh decompiler~\cite{reincarnate}
  to synthesize smaller CAD models.

\item A large scale evaluation demonstrating
  the performance and scalability of \tool
  on models downloaded from a popular online
    repository~\cite{thing}.

%% \item An empirical evaluation of \tool's
%%     performance and correctness.
%%   We compiled \thingBenchCount
%%     high-level CAD models
%%     from Thingiverse~\cite{thing}
%%     and used \tool to decompile them.
%%   \tool can quickly shrink CSG expressions,
%%     reducing the size of inputs
%%     compiled from large models by
%%     \thingBenchAvgImprove on average.

\end{itemize}

This paper proceeds gradually,
  first introducing \caddy and a
  running example (\autoref{sec:overview}).
\tool primarily exploits opportunities
  to ``reroll loops'' (\autoref{sec:loops}).
Finding such opportunities
  is challenging due to variations in
  mesh decompiler output \chOne,
so \tool uses \egraphs to
  implement a robust
  CAD rewrite system (\autoref{sec:egraphs}).
Finding the right CAD reordering is
  crucial to expose high-level structure \chTwo,
  but difficult with rewrites alone
  due to AC-matching.
Solvers in \tool propagate profitable reorderings
  through the \egraph
  by unifying order-inequivalent
  expressions annotated with
  \semts (\autoref{sec:inverse}).

We developed a library of
  \toolNumRules CAD rewrites
  and prototyped \tool in
  \toolLinesRust lines of Rust
  (\autoref{sec:impl}).
\autoref{sec:eval} shows how composing
  \tool with an existing mesh decompiler~\cite{reincarnate}
  qualitatively improves editability
  (sketched in \autoref{fig:motiv})
  and describes an evaluation of
  \tool's performance and correctness
  on real-world CAD models
  downloaded from Thingiverse.
\autoref{sec:related} briefly surveys the
  most relevant related work and
  \autoref{sec:conclusion} concludes.

\section{\caddy and Second Stage Decompilation}
\label{sec:overview}
\label{sec:caddy}

The \emph{\caddy} language (\autoref{fig:grammar1}) provides
  map- and fold-like functional list operators to express
  repetitive structure in CAD models,
  as well as a \emph{Core \caddy} fragment that
  corresponds directly to CSG.
The \caddy semantics
  fully unroll a program's functional list operators
  to produce a Core \caddy (CSG) expression.
\tool ``goes the other way,''
  decompiling a Core Caddy expression to
  a Caddy program that aims to
  expose latent repetitive structure.
This section introduces a running example
  that subsequent sections extend
  to illustrate challenges that arise when
  shrinking noisy, unstructured outputs
  from existing mesh decompilers.

\subsection{Core Caddy, Caddy, Equivalence}
\label{subsec:caddylang}

Core \caddy includes various primitives parametrized by dimensions---
  cuboids parametrized by side length,
  spheres by radius,
  cylinders and hexagonal prisms by height and radius,
  \etc.
\caddy also provides binary\footnote{
    We use syntactic sugar to present binary nested operators as
    left-associative over multiple arguments, e.g.,
    \F{(Union a b c)} means \F{(Union (Union a b) c)}.}
  set theoretic operators
  \F{Union}, \F{Difference}, and \F{Intersection}, and
affine\footnotemark{} transformations
  like \F{Translate}, \F{Rotate}, and \F{Scale} that
  are parameterized by 3D vectors.
For example, \F{(Translate [1,0,0] (Sphere 2))}
  shifts a sphere with radius 2
  a single unit of distance along the x-axis.
\F{TranslateSpherical} (not present in Core Caddy or CSG)
  captures a common pattern in models relying on translations in
  spherical rather than Cartesian coordinates.
\footnotetext{
  Here affine means that parallel lines remain
  parallel after transformation.
}

% cannot use mathbb or math superscripts in grammar environment
\newcommand{\reals}{ℝ}
\newcommand{\posints}{ℤ\textsuperscript{+}}

\begin{figure}
  \sf

  \setlength{\grammarparsep}{5pt plus 1pt minus 1pt}
  \setlength{\grammarindent}{4.5em}
  \begin{grammar}
    <op> ::= + | - | $\times$ | /
    \quad \quad
    % do not use <num> on LHS of ::=
    % grammar will format inconsistently w/ other classes
    num \, ::= \, \reals\ | <var> | <num> <op> <num>

    <vec2> ::= [<num>, <num>]
    \quad \quad
    % do not use <vec3> on LHS of ::=
    % grammar will format inconsistently w/ other classes
    vec3 \, ::= \, [<num>, <num>, <num>]

    <affine> ::= Translate | Rotate | Scale | {\bf TranslateSpherical}

    <binop> ::= Union | Difference | Intersection

    <cad> ::= (Cuboid <vec3>) \;|\; (Sphere <num>) \;\;
    \alt (Cylinder <vec2>) \;|\; (HexPrism <vec2>) \;|\; \ldots
    \alt (<affine> <vec3> <cad>)
    \alt (<binop> <cad> <cad>)
    \alt {\bf(Fold} <binop> <cad-list>{\bf)}

    <cad-list> ::= (List <cad>+)
    \alt (Concat <cad-list>+)
    \alt (Tabulate (<var> \, \posints)+ <cad>)
    %\alt (Repeat \posints\ <cad>) \todo{repeat just sugar?}
    \alt (Map2 <affine> <vec3-list> <cad-list>)

    <vec3-list> ::= (List <vec3>+)
    \alt (Concat <vec3-list>+)
    \alt (Tabulate (<var> \, \posints)+ <vec3>)
    %\alt (Repeat \posints\ <vec3>) \todo{repeat just sugar?}

  \end{grammar}
 \caption{
   \caddy syntax.
   The Core \caddy (CSG) subset omits variables,
    list forms (those using \textbf{\fold}),
    and \textbf{\F{TranslateSpherical}}.
 }
 \label{fig:grammar1}
\end{figure}

%%% Local Variables:
%%% TeX-master: "main"
%%% End:

\autoref{fig:sem1} gives semantics
  for the functional list operators \caddy
  provides on top of Core \caddy.
\mapi takes pairs of variables and positive integers
  $(x_1 \, b_1) \,...\, (x_n \, b_n)$
  as well as a \caddy expression $e$, and
  returns the list of length $\Pi b_i$ generated by
  $n$ nested loops evaluating $e$
  over the variables $x_1 \,...\, x_n$ up to
  the bounds $b_1 \,...\, b_n$:
  $$
    (\slist \;\; \subst{e}{x_1}{0}...\subst{}{x_n}{0} \;\; \ldots \;\;
    \subst{e}{x_1}{b_1 - 1}...\subst{}{x_n}{b_n - 1})
  $$
  where \subst{e}{x}{i} denotes
  substituting all free occurrences
  (not bound by nested \mapis) of
  $x$ in $e$ with $i$.
For example,
\begin{lstlisting}[xleftmargin=0em]
(Tabulate (i 2) (j 3) (Cuboid [2 $\times$ i + 2, 7, j + 1])) $\Rightarrow$
    (List (Cuboid [2, 7, 1]) (Cuboid [2, 7, 2]) (Cuboid [2, 7, 3])
          (Cuboid [4, 7, 1]) (Cuboid [4, 7, 2]) (Cuboid [4, 7, 3]))
\end{lstlisting}
For the frequent special case of
  \F{(\mapi ($x$ $n$) $e$)} when
  $x$ is not free in $e$,
  we write \F{(\rpt $n$ $e$)} as syntactic sugar.
%  for generating the list of $n$ copies of $v$ where $e \Rightarrow v$.

%% \todo{ztatlock: do we need the following restriction? I suspect not?}
%% Substitution for \mapi happens all at once; in
%%   other words, the bindings cannot be nested.
%% If a \mapi is nested within a \mapi, all loop variables within the
%%   inner \mapi refer to the inner \mapi, not the outer.

%% On top of Core Caddy,
%%   Caddy provides functional list operators
%%   to express repetitive structure.
%% \tool can evaluate Caddy programs down to Core Caddy expressions
%%   following the semantics in \autoref{fig:sem1}.
%% Most important among these constructs is \mapi, which takes three
%%   integer loop bounds\footnotemark{} and a \caddy expression,
%%   returning a list of expressions generated by three nested loops over
%%   the variables $i,j,k$ up to the three loop bounds.
%% For example, the \caddy expression
%%   \lstinline[mathescape]{(Tabulate 2 3 1 (Cuboid [$2i + 2$, 7, $j + 1$]))}
%%   evaluates to:
%% \begin{lstlisting}[xleftmargin=2em]
%% (List (Cuboid [2, 7, 1]) (Cuboid [2, 7, 2]) (Cuboid [2, 7, 3])
%%       (Cuboid [4, 7, 1]) (Cuboid [4, 7, 1]) (Cuboid [4, 7, 1]))
%% \end{lstlisting}

%% \rpt and \concat provide semantics related to their functional
%%   programming counterparts.

\map produces a list of Core \caddy expressions by
  applying an affine operator to a
  list of transformation parameters and a
  list of CAD arguments.
For example,
%  \F{(Map2 Scale (List [2,2,2] [3,3,3]) (Repeat 2 (Sphere 1)))}
\begin{lstlisting}[xleftmargin=1em]
(Map2 Scale (List [2,2,2] [3,3,3]) (Repeat 2 (Sphere 1))) $\Rightarrow$
    (List (Scale [2,2,2] (Sphere 1)) (Scale [3,3,3] (Sphere 1)))
\end{lstlisting}

\begin{figure}
  \centering
  \footnotesize

  \newcommand\itrm[1]{\rm \it #1}
  \newcommand\semrulesep{1.75em}

  \begin{equation*}
  \hspace*{-1.5em}
  \begin{array}{c}
%%  \inference[]
%%  {e \Rightarrow v}
%%  {\F{(Repeat $n$ $e$)} \Rightarrow \F{(List $v$...\; \text{$n$ times})}}
%%  \vspace{2em}\\

  \inference[]
    { e \Rightarrow \F{(List $v_1$ ... $v_n$)} \quad
    \F{$f_1$ = $v_1$} \quad
    \F{$f_i$ = $(\mathit{binop} \, f_{i-1} \, v_i)$} }
  { \F{(Fold $\mathit{binop}$ $e$)} \Rightarrow \F{$f_n$} }
  \\[\semrulesep]

  \inference[]
  {e \Rightarrow \F{
      (List
        (List $v_{1,1}$ $v_{1,2}$ ...)
        (List $v_{2,1}$ $v_{2,2}$ ...) ...)}}
  {\F{(Concat $e$)} \Rightarrow \F{(List $v_{1,1}$ $v_{1,2}$ ... $v_{2,1}$ $v_{2,2}$ ...)}}
  \\[\semrulesep]

  \inference[]
  {\subst{e}{x_1}{i_1}...\subst{}{x_n}{i_n} \Rightarrow v_{(i_1, \, ... \, , i_n)}}
  {\F{(\mapi $(x_1 \, b_1)$ ... $(x_n \, b_n)$ $e$)}
   \Rightarrow
    \F{(List $v_{(0, \, ... \, , 0)}$ \, ... \, $v_{(b_1 - 1, \, ... \ , b_n - 1)}$ )}
  }
  \\[\semrulesep]

%%  \inference[]
%%  {e[(i, j, k)] \Rightarrow v_{ijk}}
%%  {\F{(MapI $b_i$ $b_j$ $b_k$ $e$)}
%%   \Rightarrow
%%   \F{(List $v_{ijk}$ ...
%%       $\forall i \in [b_i], j \in [b_j], k \in [b_k]$)}
%%  }
%%  \vspace{1mm} \\
%%  \textrm{Elements $v_{ijk}$ are ordered lexicographically by tuple $(i, j, k)$}
%%  \\[\semrulesep]

  \inference[]
  {
  \mathit{ps} \Rightarrow \F{(List [$a_1$,$b_1$,$c_1$] [$a_2$,$b_2$,$c_2$] ...)} \qquad
   \mathit{es} \Rightarrow \F{(List $v_1$ $v_2$ ...)}
  }
    {\F{(Map2 $\mathit{affine}$ $\mathit{ps}$ $\mathit{es}$)} \Rightarrow
      \F{(List \,
        $(\textit{affine} \, [a_1,b_1,c_1] \, v_1)$ \,
        $(\textit{affine} \, [a_2,b_2,c_2] \, v_2)$ ...)}}
  \\[\semrulesep]

%%  \inference[]
%%  {
%%  \mathit{params} \Rightarrow \F{(List [$a_1$,$b_1$,$c_1$] [$a_2$,$b_2$,$c_2$] ...)} \\
%%   \mathit{es} \Rightarrow \F{(List $v_1$ $v_2$ ...)} \quad
%%    \textit{aff}_i = \F{($\mathit{affine}$ [$a_i$,$b_i$,$c_i$] $v_i$)} \\
%%  }
%%    {\F{(Map2 $\mathit{affine}$ $\mathit{params}$ $\mathit{es}$)} \Rightarrow \F{(List $\textit{aff}_1$ $\textit{aff}_2$ ...)}}
%%  \\[\semrulesep]

  \inference[]
  {e \Rightarrow v \qquad
  \textsf{to\_cartesian}(r,\phi,\theta) = (x,y,z)
  }
  {\F{(TranslateSpherical $[r,\phi,\theta]$ $e$)}
   \Rightarrow
   \F{(Translate $[x,y,z]$ $v$)}
  }
  \end{array}
  \end{equation*}

  \caption{
    Big step semantics reducing
      well-formed \caddy programs to
      Core \caddy expressions.
    \subst{e}{x}{i} denotes substituting
      all free occurrences of $x$ in $e$ with $i$.
    Additional rules (not shown) also
      evaluate under \F{List}, affines, and binops.
    %%Well-formed \caddy programs evaluate to Core \caddy programs.
    %%$e[(i,j,k)]$ denotes substituting the integers $i,j,k$ for the
    %%list variables \F{i, j, k} in expression $e$.
    %%$[b_i]$ denotes the sequence of natural numbers up to $b_i$, exclusive.
  }
  \label{fig:sem1}
\end{figure}

%%% Local Variables:
%%% TeX-master: "main"
%%% End:

\caddy programs are equivalent iff they evaluate to
  equivalent Core \caddy programs.
By design, Core \caddy directly corresponds to CSG,
  whose semantics is given in prior work \cite{ronse, reincarnate, sherman19}.
\autoref{sec:eval} describes practically testing \caddy equivalence by
  evaluating programs to Core \caddy, compiling them to meshes,
  and comparing Hausdorff distances.\footnote{Informally,
    the Hausdorff distance between two meshes is small if
    every point on each mesh is near some point on the other.}
%%  between meshes with a small epsilon to
%%  account for floating point rounding error.

%% \footnotetext{
%%   Syntactically, \mapi always takes three loop bounds, but we
%%   elide inner loops with bound 1.
%% }

\subsection{A Running Example for Shrinking Caddy}
\label{subsec:prob}

% \captionsetup[sub]{font=footnotesize,labelfont={bf,sf},aboveskip=0pt,belowskip=-4pt}

\begin{figure}
  \begin{subfigure}[b]{0.4\linewidth}
    \centering
    \includegraphics[height=28mm]{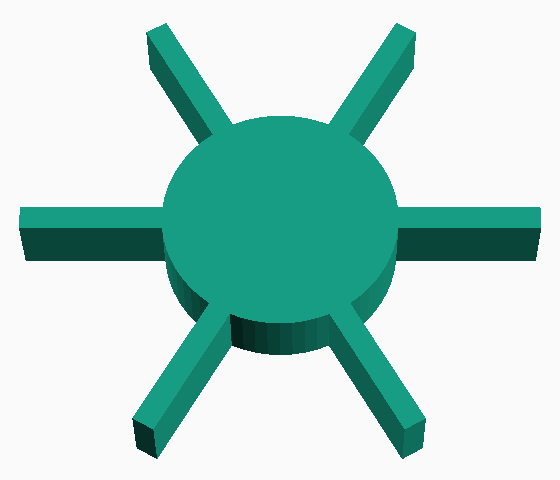}
    \hfill
    \caption{CAD model of ship's wheel}
    \label{fig:ship-cad}
  \end{subfigure}
  \hfill
  \begin{subfigure}[b]{0.5\linewidth}
    \begin{lstlisting}
(Union
 (Cylinder [1, 5, 5])
 (Fold Union
  (Tabulate (i 6)
   (Rotate $[0,0,60i]$
    (Translate $[1,-0.5, 0]$
     (Cuboid $[10,1,1]$))))))
    \end{lstlisting}
    \caption{\caddy program}
    \label{fig:ship-prog}
  \end{subfigure}

  \begin{subfigure}{\linewidth}
    \begin{lstlisting}[basicstyle=\scriptsize\sffamily]
(Union
 (Cylinder |$[1,5]$|)
 (Union
  (Rotate |$[0,0,0]$|   (Translate |$[1,-0.5,0]$| (Cuboid |$[10,1,1]$|)))
  (Rotate |$[0,0,60]$|  (Translate |$[1,-0.5,0]$| (Cuboid |$[10,1,1]$|)))
  (Rotate |$[0,0,120]$| (Translate |$[1,-0.5,0]$| (Cuboid |$[10,1,1]$|)))
  (Rotate |$[0,0,180]$| (Translate |$[1,-0.5,0]$| (Cuboid |$[10,1,1]$|)))
  (Rotate |$[0,0,240]$| (Translate |$[1,-0.5,0]$| (Cuboid |$[10,1,1]$|)))
  (Rotate |$[0,0,300]$| (Translate |$[1,-0.5,0]$| (Cuboid |$[10,1,1]$|)))))
    \end{lstlisting}
    \caption{
      Ideal Core \caddy expression that exposes structure
    }
    \label{fig:ship-csg-good}
  \end{subfigure}
  \begin{subfigure}{\linewidth}
    \begin{lstlisting}[basicstyle=\scriptsize\sffamily]
(Union
  (Rotate |$[0,0,120]$| (Translate |$[1,-0.5,0]$| (Cuboid |$[10,1,1]$|)))
  (Scale |$[10,1,1]$| (Translate |$[0.1,-0.5,1]$| (Cuboid |$[1,1,1]$|)))
  (Rotate |$[0,0,300]$| (Translate |$[1,-0.5,0]$| (Cuboid |$[10,1,1]$|)))
  (Scale |$[5,5,1]$| (Cylinder |$[1,1]$|))
  (Translate |$[-1,0.5,0]$| (Scale |$[-1,-1,1]$| Cuboid |$[10,1,1]$|))
  (Rotate |$[0,0,240]$| (Translate |$[1,-0.5,0]$| (Cuboid |$[10,1,1]$|)))
  (Rotate |$[0,0,60]$| (Translate |$[1,-0.5,0]$| (Cuboid |$[10,1,1]$|))))
    \end{lstlisting}
    \caption{
      Equivalent Core \caddy expression that obfuscates structure
    }
    \label{fig:ship-csg-bad}
  \end{subfigure}

  \caption{
    (\subref{fig:ship-cad}) CAD model for a ship's wheel.
    (\subref{fig:ship-prog}) \caddy features like \mapi express
      repeated design components.
    Such repetition can be obvious in
      Core Caddy (\subref{fig:ship-csg-good}),
      but existing mesh decompilers
      obfuscate structure (\subref{fig:ship-csg-bad}).
  }
  \label{fig:ship}
\end{figure}

%Even this simple model illustrates the challenge of
%  shrinking a Core \caddy (CSG) expression into
%  a parameterized, easier-to-edit \caddy program.

\autoref{fig:ship-cad} shows a
  simple CAD model of a ship's wheel and
  \autoref{fig:ship-prog} shows the
  corresponding desired \caddy output from \tool.
\autoref{fig:ship-prog} reifies repetitive structure:
  making a change to all the spokes
  only requires a single edit instead of
  six coordinated modifications in different locations.

When repetitive structure is easily exposed,
  as in the ideal Core \caddy of \autoref{fig:ship-csg-good},
%% Ideally, repetitive structure would
%%   always be evident in the Core \caddy expression
%%   \tool takes as input (\autoref{fig:ship-csg-good}).
%% In such cases,
%%   repetition is easily detected and
  solvers can infer the
  arithmetic function relating instances
  of repeated design components.
\autoref{sec:loops} describes
  \tool's rewrite-driven approach to
  infer such functions and shrink programs
  by rerolling loops.
%%  exposing repetitive structure using \map,
%%  using simple solvers to replace lists of parameters with \mapis, and
%%  finally employing \mapi again to
%%  parameterize repeated components.

In practice,
  given a mesh representing \autoref{fig:ship-cad},
  mesh decompilers can generate CSG expressions
  equivalent to \autoref{fig:ship-csg-good}, but
  which obfuscate repetitive structure.
%  as in \autoref{fig:ship-csg-bad}.
Affine transformations may be different or missing and,
  from a solver's perspective,
  lists may be inconveniently ordered or partitioned.
Comparing Figure~\ref{fig:ship-csg-good} to \ref{fig:ship-csg-bad},
  \F{Rotate [0,0,180]} has been replaced with
  an equivalent \F{Scale [-1,-1,1]},
  identity transformations have been omitted,
  the \F{Union} has been reordered,
  and \F{Scales} and \F{Translates} have been inconsistently swapped.
Sections \ref{sec:egraphs} and \ref{sec:inverse}
  walk through progressively more challenging variants of
  Core \caddy inputs for the ship's wheel to illustrate
  how \tool uses \egraphs and \semts
  to robustly handle such variation.

\section{Shrinking \caddy by Rerolling Loops}
\label{sec:loops}
\label{subsec:other}

\tool shrinks repetitive \caddy programs
  by ``rerolling loops''.
%  by exposing the repetition in \slists
%  that it rewrites into generalized \mapis
%  to ``reroll loops''.
First, rewrites \textit{find structure} by
  separating affine operators from
  their parameters and CAD arguments under \maps.
This can expose program repetition as
  repetitive \slists.
%  \footnote{
%    When combined with CAD rewrites from \autoref{sec:egraphs},
%    this separation also factors out CAD primitive parameters.}
Next, arithmetic solvers find
  equivalent closed form \mapis
  for repetitive lists.
These \mapis generalize the program
  and provide parameters that simplify future edits.
%  (e.g., sequences of 3D positions with consistent radial spacing).
Finally, rewrites \textit{restore structure}
  by recombining the (generalized)
  affine parameters and CAD arguments
  from \maps into a single \mapi.
\autoref{fig:loop-rules} shows this strategy's key rewrites.

%% \footnote{
%%   The \textbf{\F{Binop Fold}} ``meta rule''
%%   pulls nested left-associated binops and
%%   right-associated \union and \inter operators into \slists.
%%   \diff is neither associative nor commutative.}

Because \tool uses an \egraph,
  these rewrites can actually be repeatedly applied in any order
  and still efficiently yield the same final result.
For simplicity,
  this section steps through the ship's wheel
  example assuming a particular fortuitous order of rewrites
  that just so happens to nicely shrink the
  ideal Core \caddy input from \autoref{fig:ship-csg-good}.

%%  these rewrites are iteratively applied
%%  along with additional CAD identities
%%  in an \egraph (\autoref{sec:egraphs}) and
%%  interact with \semts (\autoref{sec:inverse}).

%% The primary way \tool eliminates repetitive structure is by finding
%%   lists with identical CAD structure and
%%   rewriting their parameter list into an equivalent \mapi.
%% First, \tool separates CAD structure from parameters using \maptwo.
%% Then, an arithmetic solver tries to rewrite the parameter list
%%   to a \mapi, and finally \tool can ``reroll'' the loop by folding
%%   the CAD structure back under the \mapi.
%% The rewrite rules used for this are shown in
%%   \autoref{fig:loop-rules}.
%% \todo{eva: are rules applied sequentially?}

\subsection{Finding Structure: A Bird's-eye View}
\label{subsec:struct}

\begin{figure}
\begin{tabular}{l}
\Rewritename{Binop Fold} \\
    \begin{lstlisting}[style=rewrite]
    ($\mathit{binop}$ $\mathit{c}_1$ $\mathit{c}_2$ ...)
    \end{lstlisting}
\; \rewritesto \;
    \begin{lstlisting}[style=rewrite]
    (Fold $\mathit{binop} \,$ (List $\mathit{c}_1$ $\mathit{c}_2$ ...))
    \end{lstlisting}
\rewritespace
\\
\Rewritename{Structure Finding} \\
    \begin{lstlisting}[style=rewrite]
    (List ($\mathit{aff}$ $\mathit{p}_1$ $\mathit{c}_1$) ($\mathit{aff}$ $\mathit{p}_2$ $\mathit{c}_2$) ...)
    \end{lstlisting}
\; \rewritesto \;
    \begin{lstlisting}[style=rewrite]
    (Map2 $\mathit{aff}$ (List $\mathit{p}_1$ $\mathit{p}_2$ ...) (List $\mathit{c}_1$ $\mathit{c}_2$ ...))
    \end{lstlisting}
\rewritespace
\\
\Rewritename{Repeat} \\
    \begin{lstlisting}[style=rewrite]
    (List $a$ $a$ $a$ ... $n\; \textrm{times}$)
    \end{lstlisting}
\; \rewritesto \;
    \begin{lstlisting}[style=rewrite]
    (Repeat $n$ $a$)
    \end{lstlisting}
\rewritespace
\\
\Rewritename{List Solve (single loop)} \\
    \begin{lstlisting}[style=rewrite]
    (List $[f_x(0),f_y(0),f_z(0)]$ $\;$ ... $\;$  $[f_x(n - 1),f_y(n - 1),f_z(n - 1)]$)
    \end{lstlisting} \\[-0.2em]
\;\;\; \rewritesto \;
    \begin{lstlisting}[style=rewrite]
    (Tabulate $(i ~ n) \;$ $[f_x(i),f_y(i),f_z(i)]$)
    \end{lstlisting}
\rewritespace
\\
\Rewritename{Repeat over Map2} \\
    \begin{lstlisting}[style=rewrite]
    (Map2 $\mathit{aff}$ (Repeat $n$ $\mathit{p}$) (Repeat $n$ $\mathit{c}$)
    \end{lstlisting}
\; \rewritesto \;
    \begin{lstlisting}[style=rewrite]
    (Repeat $n$ ($\mathit{aff}$ $\mathit{p}$ $\mathit{c}$))
    \end{lstlisting}
\rewritespace
\\
\Rewritenamecond{Tabulate over Map2}{where $b = \Pi b_i$} \\
    \begin{lstlisting}[style=rewrite]
    (Map2 $\mathit{aff}$ $\;$ (Tabulate $(x_1 ~ b_1)$ ... $\mathit{p}$) $\;$ (Tabulate $(x_1 ~ b_1)$ ... $\mathit{c}$))
    \end{lstlisting} \\[-0.2em]
\;\;\; \rewritesto \;
    \begin{lstlisting}[style=rewrite]
    (Tabulate $(x_1 ~ b_1)$ ...  ($\mathit{aff}$ $\mathit{p}$ $\mathit{c}$))
    \end{lstlisting}
    \\
    \begin{lstlisting}[style=rewrite]
    (Map2 $\mathit{aff}$ (Tabulate $(x_1 ~ b_1)$ ... $\mathit{p}$) (Repeat $b$ $\mathit{c}$))
    \end{lstlisting}
\; \rewritesto \;
    \begin{lstlisting}[style=rewrite]
    (Tabulate $(x_1 ~ b_1)$ ...  ($\mathit{aff}$ $\mathit{p}$ $\mathit{c}$))
    \end{lstlisting}
    \\
    \begin{lstlisting}[style=rewrite]
    (Map2 $\mathit{aff}$ (Repeat $b$ $\mathit{p}$) (Tabulate $(x_1 ~ b_1)$ ... $\mathit{c}$))
    \end{lstlisting}
\; \rewritesto \;
    \begin{lstlisting}[style=rewrite]
    (Tabulate $(x_1 ~ b_1)$ ...  ($\mathit{aff}$ $\mathit{p}$ $\mathit{c}$))
    \end{lstlisting}
\end{tabular}
\caption{Rewrite rules for loop rerolling}
\label{fig:loop-rules}
\end{figure}

%%% Local Variables:
%%% TeX-master: "main"
%%% End:

Applying \textbf{\F{Binop Fold}} to the
inner \union in \autoref{fig:ship-csg-good} produces:

{\footnotesize
\noindent
\begin{minipage}[t]{0.4\linewidth}
    \begin{lstlisting}[style=rewrite]
    (Union (Union (Union ...
     (Rotate |$[0,0,0]$|   $\mathit{cad}_1$)
     (Rotate |$[0,0,60]$|  $\mathit{cad}_2$)
     (Rotate |$[0,0,120]$| $\mathit{cad}_3$) ...)
    \end{lstlisting}
\end{minipage}
\begin{minipage}[t]{0.1\linewidth}
  \begin{align*}
  \\[-5pt]
  \blacktriangleright
  \end{align*}
\end{minipage}
\begin{minipage}[t]{0.4\linewidth}
    \begin{lstlisting}[style=rewrite]
    (Fold Union (List
     (Rotate |$[0,0,0]$|   $\mathit{cad}_1$)
     (Rotate |$[0,0,60]$|  $\mathit{cad}_2$)
     (Rotate |$[0,0,120]$| $\mathit{cad}_3$) ...))
    \end{lstlisting}
\end{minipage}
}

\vspace{0.1in}
A structure finder
  (detailed in \autoref{sec:egraphs})
  searches for a list of affine transformations
  all using the same operator $\mathit{aff}$.
\textbf{\F{Structure Finding}} separates the
  affine parameters and CAD arguments out into
  two \slists under a \map with $\mathit{aff}$:

{\footnotesize
\noindent
\begin{minipage}[t]{0.4\linewidth}
    \begin{lstlisting}[style=rewrite]
    (Fold Union (List
     (Rotate |$[0,0,0]$|   $\mathit{cad}_1$)
     (Rotate |$[0,0,60]$|  $\mathit{cad}_2$)
     (Rotate |$[0,0,120]$| $\mathit{cad}_3$) ...))
    \end{lstlisting}
\end{minipage}
\begin{minipage}[t]{0.1\linewidth}
  \begin{align*}
  \\[-5pt]
  \blacktriangleright
  \end{align*}
\end{minipage}
\begin{minipage}[t]{0.45\linewidth}
    \begin{lstlisting}[style=rewrite]
    (Fold Union
     (Map2 Rotate
      (List $[0,0,0]$ $[0,0,60]$ $[0,0,120]$ ...)
      (List $\mathit{cad}_1$ $\mathit{cad}_2$ $\mathit{cad}_3$ ...)))
    \end{lstlisting}
\end{minipage}
}

\vspace{0.1in}
The structure finder is applied repeatedly.
Here it exposes lists of identical elements,
  letting the \textbf{\F{Repeat}} rewrite produce:

{\footnotesize
\noindent
\begin{minipage}[t]{0.45\linewidth}
\begin{lstlisting}[basicstyle=\scriptsize\sffamily]
(Fold Union
 (Map2 Rotate
  (List $[0,0,0]$ $[0,0,60]$ $[0,0,120]$ ...)
  (Map2 Translate
   (List $[1,-0.5,0]$ ...)
   (List (Cube $[10,1,1]$) ...))))
\end{lstlisting}
\end{minipage}
\begin{minipage}[t]{0.1\linewidth}
  \begin{align*}
    \\[2pt]
  \blacktriangleright
  \end{align*}
\end{minipage}
\begin{minipage}[t]{0.45\linewidth}
\begin{lstlisting}[basicstyle=\scriptsize\sffamily]
(Fold Union
 (Map2 Rotate
  (List $[0,0,0]$ $[0,0,60]$ $[0,0,120]$ ...)
  (Map2 Translate
   (Repeat 6 $[1,-0.5,0]$)
   (Repeat 6 (Cube $[10,1,1]$)))))
\end{lstlisting}
\end{minipage}
}

\subsection{Introducing \mapi by Solving Lists}
\label{subsec:solve}

Once structure finding has
  isolated a \slist of vectors $\ell$,
  arithmetic solvers attempt to find
  equivalent \mapis.
The current \tool prototype
  provides simple solvers for
  first- and second-degree polynomials in both
  Cartesian and spherical coordinates.
Given $\ell = (\slist ~ [x_1, y_1, z_1] ~ ... ~ [x_n, y_n, z_n])$,
  these solvers infer independent functions
  $f_x$, $f_y$, $f_z$ for the
  $x$, $y$, $z$ components of $\ell$ respectively.
%% The \textbf{\F{List Solve}} rewrite can then apply:
%%   $\ell \blacktriangleright$
%%   \F{(Tabulate $(i ~ n) \;$ $[f_x(i),f_y(i),f_z(i)]$)}.
In practice,
  running arithmetic solvers on
  floating point numbers output by
  existing mesh decompilers requires
  accepting \mapis within some $\epsilon$ of $\ell$,
  especially for tools that rely on
  randomized algorithms~\cite{inverse}
  like RANSAC~\cite{ransac}.
%\todo{ztatlock: massage wording about solvers, say a bit more, or point to details in impl?}

For the \F{Rotate} parameters \F{(\slist \, $[0,0,0]$ \, $[0,0,60]$ \, ... \, $[0,0,300]$)},
  solvers find \F{(Tabulate $(i ~ 6) \;$ $[0, 0, 60i]$)}.
\textbf{\F{List Solve}} then produces:

{\footnotesize
\noindent
\begin{minipage}[t]{0.45\linewidth}
\begin{lstlisting}[basicstyle=\scriptsize\sffamily]
(Fold Union
 (Map2 Rotate
  (List $[0,0,0]$ $[0,0,60]$ $[0,0,120]$ ...)
  (Map2 Translate
   (Repeat 6 $[1,-0.5,0]$)
   (Repeat 6 (Cube $[10,1,1]$)))))
\end{lstlisting}
\end{minipage}
\begin{minipage}[t]{0.1\linewidth}
  \begin{align*}
    \\[2pt]
  \blacktriangleright
  \end{align*}
\end{minipage}
\begin{minipage}[t]{0.45\linewidth}
\begin{lstlisting}[basicstyle=\scriptsize\sffamily]
(Fold Union
 (Map2 Rotate
  (Tabulate $(i ~ 6) \;$ $[0, 0, 60i]$)
  (Map2 Translate
   (Repeat 6 $[1,-0.5,0]$)
   (Repeat 6 (Cube $[10,1,1]$)))))
\end{lstlisting}
\end{minipage}
}

\vspace{0.1in}
In this example,
  the solvers relied on their input
  arriving in just the right order.
\autoref{sec:inverse} shows how \semts allow
  solvers to \textit{reorder} their input
  to infer better \mapis while preserving equivalence.

\subsection{The Final Squeeze: Recombining \maps}
\label{subsec:restore}

Finally,
  since both the \rpts and \mapi have matching bounds,
  \textbf{\F{Repeat over Map2}} and
  \textbf{\F{Tabulate over Map2}} recombine the
  separated affine parameters and CAD arguments
  to produce the desired output from
  the inner \union of \autoref{fig:ship-csg-good}:

{\footnotesize
\noindent
\begin{minipage}[t]{0.4\linewidth}
    \begin{lstlisting}[style=rewrite]
    (Fold Union
     (Map2 Rotate
      (Tabulate (i 6) $[0,0,60i]$)
       (Map2 Translate
        (Repeat 6 $[1,-0.5,0]$)
        (Repeat 6 (Cube $[10,1,1]$)))))
    \end{lstlisting}
\end{minipage}
\begin{minipage}[t]{0.1\linewidth}
  \begin{align*}
  \\[-5pt]
  \blacktriangleright
  \end{align*}
\end{minipage}
\begin{minipage}[t]{0.45\linewidth}
    \begin{lstlisting}[style=rewrite]
    (Fold Union
     (Tabulate (i 6)
      (Rotate $[0,0,60i]$
       (Translate $[1,-0.5,0]$
        (Cube $[10,1,1]$)))))
    \end{lstlisting}
\end{minipage}
}

\vspace{0.1in}
This section illustrated \tool's core strategy:
  shrinking \caddy by rerolling loops.
However, the example relied on a
  specific rewrite order and
  \autoref{fig:ship-csg-good} as an
  unrealistically ideal input.
Subsequent sections show how
  \egraphs and \semts enable
  \tool to robustly shrink
  noisy and unstructured CSGs.

%% In the ship's wheel example, once the structure finder
%%   isolates the list of parameters, the arithmetic solver can rewrite
%%   that list to a \mapi{}.
%% Then, the \F{Repeat over Map2} and \F{Tabulate over Map2} rules
%%   combine the separated CAD structure with the inferred \mapi,
%%   yielding the desired output:
%%
%% The \F{Tabulate over Map2} rules must ensure that product of the \mapi's
%%   bounds match the length of the \rpt{} in the other position of the
%%   \map.

%%% Local Variables:
%%% TeX-master: "main"
%%% End:

\section{\egraphs and CAD Equality Saturation}
\label{sec:egraphs}

%\todo{ztatlock: \tool easy to extend, just add rewrites}

\label{sec:appr}
Rewrites to shrink \caddy by rerolling loops
  must be applied in just the right order
  to programs that already make structure apparent
  as in \autoref{fig:ship-csg-good}.
Simply interleaving additional CAD rewrites to
  expose repetitive structure
%  and exploring many potential orderings
  initially seems infeasible because
  the necessary rewrites are not confluent
%  \footnote{
%    \todo{ztatlock: footnote on confluence and why we don't have it}
%  }
  and the space of possible orderings explodes exponentially.
However, past work on Equality Saturation~\cite{eqsat}
  demonstrates how \egraphs~\cite{nelson} can make this strategy
  efficient for many rewrite rules.
This section shows how \tool applies
  Equality Saturation in the CAD domain
  to robustly handle CSG variations
  when shrinking \caddy programs.

%% \tool uses \egraphs to implement its rewrite-driven shrinking of flat CSG
%%   programs.
%% \egraphs allow the loop rerolling discussed above to work on \mtoc
%%   outputs where the repetitive structure is only visible under a set
%%   of CAD identities.
%% This section provides necessary background on \egraphs, using them for
%%   rewrite-driven optimization, and presents some limitations related
%%   the AC-matching problem that \tool addresses.

%\todo{for zach: explain that we dont have beta reduction, explain mapi story.}

\subsection{Rewrite Phase Ordering: What, When, Where} % Motivation for \egraphs
\label{subsec:eg-motiv}

A slightly perturbed \caddy example for
  the spokes of the ship's wheel
  omits \F{Rotate $[0,0,0]$} and
  replaces \F{Rotate $[0,0,180]$} by the equivalent
  \F{Scale $[-1,-1,1]$}:

%Consider the \F{Fold Union} from the Core \caddy program for the ship's wheel, where
%  the \F{Rotate $[0,0,0]$} is omitted and the \F{Rotate $[0,0,180]$}
%  is changed to an equivalent \F{Scale $[-1,-1,1]$}:

\noindent
\begin{lstlisting}[basicstyle=\scriptsize\sffamily]
(Fold Union  (List
  (Translate |$[1,-0.5,0]$| (Cube |$[10,1,1]$|))
  (Rotate |$[0,0,60]$|   (Translate |$[1,-0.5,0]$| (Cube |$[10,1,1]$|)))
  (Rotate |$[0,0,120]$|  (Translate |$[1,-0.5,0]$| (Cube |$[10,1,1]$|)))
  (Scale  |$[-1,-1,1]$| (Translate |$[1,-0.5,0]$| (Cube |$[10,1,1]$|)))
  (Rotate |$[0,0,240]$|  (Translate |$[1,-0.5,0]$| (Cube |$[10,1,1]$|)))
  (Rotate |$[0,0,300]$|  (Translate |$[1,-0.5,0]$| (Cube |$[10,1,1]$|)))))
\end{lstlisting}

The three-phase loop rerolling strategy
  from \autoref{sec:loops} now breaks:
  \tool must interleave its search
  with additional CAD rewrites (\autoref{fig:cad-rules})
  to expose the repeated affine transformations
  as in~\autoref{fig:ship-csg-good}.
This \textit{phase ordering problem}~\cite{phase-ordering-decide, eqsat}
  makes it difficult to determine
  when to apply which rewrites and where.
Poor choices will only further obfuscate repetitive structure
  and no single strategy is best in general.

%% The list order and structure is the same, but the affine
%%   transformations within the list are no longer the same, so the
%%   previously discussed loop rerolling will not work.
%% \tool must consider a set of CAD-specific identities
%%   (\autoref{fig:cad-rules}) to reveal the repetitive structure.
%% However, this breaks the apparent ordering of
%%   the phases from the previous section (separate structure, infer
%%   loops, restore structure).
%% Applying a CAD identity may reveal or further obfuscate repetitive
%%   structure, depending on the surrounding program, making it no longer
%%   clear when to apply which rewrite to which part of the program.

\begin{figure}
\begin{tabular}{lcl}
\rewritename{Affine Identities} \\
    \begin{lstlisting}[style=rewrite]
    (Rotate $[0,0,180]$ $\mathit{cad}$)
    \end{lstlisting}
    & \rewritesboth &
    \begin{lstlisting}[style=rewrite]
    (Scale $[-1,-1,1]$ $\mathit{cad}$)
    \end{lstlisting}
    \\
    \begin{lstlisting}[style=rewrite]
    (Rotate $[0,0,0]$ $\mathit{cad}$)
    \end{lstlisting}
    & \rewritesboth &
    \begin{lstlisting}[style=rewrite]
    $\mathit{cad}$
    \end{lstlisting}
    \\
    \begin{lstlisting}[style=rewrite]
    (Translate $[0,0,0]$ $\mathit{cad}$)
    \end{lstlisting}
    & \rewritesboth &
    \begin{lstlisting}[style=rewrite]
    $\mathit{cad}$
    \end{lstlisting}
    \\
    \begin{lstlisting}[style=rewrite]
    (Scale $[1,1,1]$ $\mathit{cad}$)
    \end{lstlisting}
    & \rewritesboth &
    \begin{lstlisting}[style=rewrite]
    $\mathit{cad}$
    \end{lstlisting}
\rewritespace
\\
\rewritename{Affine Interchanging} \\
    \begin{lstlisting}[style=rewrite]
    (Scale $[a,b,c]$
      (Translate $[d,e,f]$ $\mathit{cad}$))
    \end{lstlisting}
    & \rewritesboth &
    \begin{lstlisting}[style=rewrite]
    (Translate $[ad,be,cf]$
      (Scale $[a,b,c]$ $\mathit{cad}$))
    \end{lstlisting}
\rewritespace
    \\
\rewritename{Affine Combination} \\
    \begin{lstlisting}[style=rewrite]
    (Scale $[a,b,c]$
      (Scale $[d,e,f]$ $\mathit{cad}$))
    \end{lstlisting}
    & \rewritesto &
    \begin{lstlisting}[style=rewrite]
    (Scale $[ad,be,cf]$ $\mathit{cad}$)
    \end{lstlisting}
    \\
    % \begin{lstlisting}[style=rewrite]
    % (Rotate $[a,b,c]$
    %   (Rotate $[d,e,f]$ $\mathit{cad}$))
    % \end{lstlisting}
    % & \rewritesto &
    % \begin{lstlisting}[style=rewrite]
    % (Rotate $[ad,be,cf]$ $\mathit{cad}$)
    % \end{lstlisting}
    % \\
    \begin{lstlisting}[style=rewrite]
    (Translate $[a,b,c]$
      (Translate $[d,e,f]$ $\mathit{cad}$))
    \end{lstlisting}
    & \rewritesto &
    \begin{lstlisting}[style=rewrite]
    (Translate $[a+d,b+e,c+f]$ $\mathit{cad}$)
    \end{lstlisting}
\rewritespace
    \\
\rewritename{Primitive-Affine Conversion} \\
    \begin{lstlisting}[style=rewrite]
    (Cuboid $[x,y,z]$)
    \end{lstlisting}
    & \rewritesboth &
    \begin{lstlisting}[style=rewrite]
    (Scale $[x,y,z]$ (Cuboid $[1,1,1]$))
    \end{lstlisting}
    \\
    \begin{lstlisting}[style=rewrite]
    (Sphere $r$)
    \end{lstlisting}
    & \rewritesboth &
    \begin{lstlisting}[style=rewrite]
    (Scale $[r,r,r]$ (Sphere 1))
    \end{lstlisting}
    \\
    \begin{lstlisting}[style=rewrite]
    (Cylinder $[h, r]$)
    \end{lstlisting}
    & \rewritesboth &
    \begin{lstlisting}[style=rewrite]
    (Scale $[r,r,h]$ (Cylinder $[1,1]$))
    \end{lstlisting}
    \\
    \begin{lstlisting}[style=rewrite]
    (Hexprism $[h, r]$)
    \end{lstlisting}
    & \rewritesboth &
    \begin{lstlisting}[style=rewrite]
    (Scale $[r,r,h]$ (Hexprism $[1,1]$))
    \end{lstlisting}
\end{tabular}
\caption{
  Selected CAD identities.
  Bidirectional arrows indicates \tool has
    a rule for each direction.
}
\label{fig:cad-rules}
\end{figure}

%%% Local Variables:
%%% TeX-master: "main"
%%% End:

Equality Saturation~\cite{eqsat} is a technique to
  mitigate phase ordering that uses
  \egraphs to compactly represent equivalence relations
  over large sets of expressions.
Instead of destructively modifying a particular concrete term,
  rewrites extend the \egraph by
  adding and unifying classes of expressions.
This eliminates the need to choose any
  particular rewrite ordering.
%  This makes \egraphs a natural fit for implementing \tool's rewrite system.
By repeatedly applying the rules in
  Figures~\ref{fig:loop-rules}~and~\ref{fig:cad-rules} to an \egraph
  and using a structure finding heuristic (\autoref{subsec:struct-eg}),
  \tool's loop rerolling strategy can robustly handle
  variations in how mesh decompilers synthesize affine operators.

%% Considering many equivalent subprograms under the CAD identities
%%   from \autoref{fig:cad-rules} means that \tool{}'s structure finding
%%   process (elaborated in \autoref{subsec:struct-eg}) can see
%%   repetitive structure not present in the original input.

\subsection{\egraph Background}
\label{subsec:eg}

\begin{figure}

\begin{minipage}{0.35\linewidth}
\begin{lstlisting}[style=rewrite]
    (Scale  |$[-1,-1,1]$|
     (Translate |$[1,-0.5,0]$|
      (Cube |$[10,1,1]$|)))
\end{lstlisting}
\end{minipage}
\hfill $\blacktriangleright$ \hfill
\begin{minipage}{0.4\linewidth}
  \begin{lstlisting}[style=rewrite]
    (Rotate |$[0,0,180]$|
     (Translate |$[1,-0.5,0]$|
      (Cube |$[10,1,1]$|)))
\end{lstlisting}
\end{minipage}

\rule{0.9\linewidth}{0.25pt}

\vspace{3mm}
\hspace*{-0.15in}
\includegraphics[width=\linewidth]{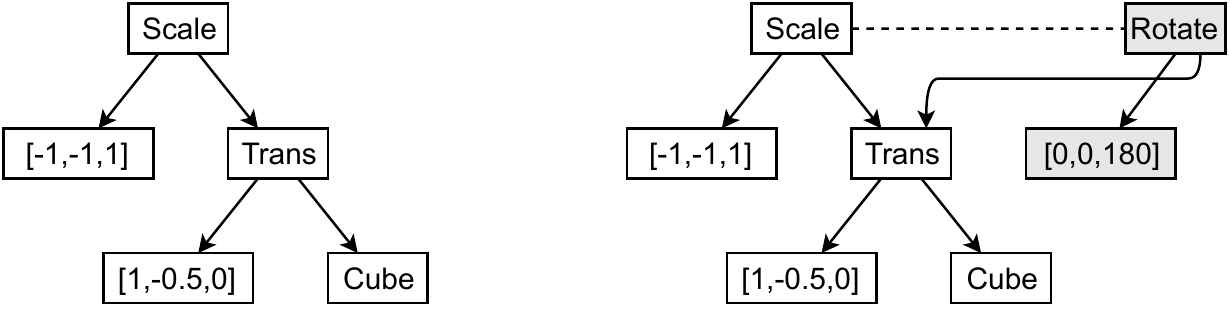}

\caption {
  \label{fig:egraph}
  \egraph before and after
    a CAD rewrite.
  Boxes represent \enodes and
    dashed edges indicate equivalence
    (membership in the same \eclass).
  Directed solid edges connect \enodes to
    their child \eclasses.
  Both the original and transformed programs
    are represented in the resulting \egraph.
}
\end{figure}

An \egraph is a set of \textit{\eclasses}, and
each \eclass is a set of equivalent \textit{\enodes}.
An \enode is an operator (\trans, \union, literal, \etc)
  applied to zero or more child \eclasses.
An \eclass $c$ \textit{represents} expression $e$ if
  $c$ contains an \enode $n$
  with the same operator as $e$ and
  the children of $n$ represent
  the children of $e$.
Each \eclass represents an exponential number of
  equivalent expressions (w.r.t. the number of \enodes),
  since each of its \enodes point to \eclasses themselves.

Adding an expression to an \egraph works bottom up:
  first add the leaves as \enodes each in their own \eclasses,
  then recursively add operators as \enodes pointing to
  the \eclasses of their operands as children.
Hashconsing ensures \enodes are
  never duplicated in an \egraph.
%% thanks to a hashconsing
%% data structure that checks if an \enode exists when adding it.
This sharing compactly represents many
  equivalent expressions.

\egraphs also provide a \emph{unify} operation that
  combines two \eclasses and
  maintains their congruence closure.
For example,
  if \eclasses $c_1$ and $c_2$ represent
  \F{(+ x y)} and \F{(+ x z)} respectively,
  then unifying the \eclasses representing \F{y} and \F{z}
  would cause $c_1$ and $c_2$ to be unified as well
  since they both contain ``+'' enodes with
  equivalent children.
\autoref{fig:egraph} shows how an
  \egraph can compactly represent equivalent expressions
  generated by rewrites, in this case,
  one of the CAD rewrites needed to expose
  repetitive structure for the ship's wheel example.

%% \autoref{fig:egraph} shows how an \egraph can compactly represent
%%   expressions proved equivalent under a set of rewrites, in this case,
%%   one of the CAD identities need to realize the structure from the
%%   ship's wheel example.
%% Boxes represent \enodes and dashed edges between them indicate that the
%% boxes are in the same \eclass.
%% Directed solid edges connect \enodes to their child \eclasses.

%\subsection{Flexible \egraph Rewrites}

\egraphs can easily be extended with
  syntactic rewrites $a \rewritesto b$:
  whenever an \eclass $c$ represents
  an expression that matches pattern $a$ under substitution $\phi$,
  the \eclass representing $\phi(b)$ is found (or constructed)
  and unified with $c$;
  the resulting eclass will represent both
  expressions $\phi(a)$ and $\phi(b)$.
Rewrites only expand the \egraph,
  all previous expressions are still represented.

We slightly generalize rewrites
  from two patterns to
  a pattern $L$ and a \textit{function} $R$ that,
  given a substitution $\phi$,
  returns an expression to be added to the
  \egraph and unified with the
  \eclass that matched $L$.
This generalization allows rewrites to implement
  rules which are not purely syntactic,
  like constant folding
  (ex: rewriting $2 + 3$ to $5$).
Many of \tool{}'s list-manipulating rewrites
  are implemented this way,
  which is convenient for rules like
  \textbf{\F{Repeat}} which need to extract
  the length of a matched list pattern.
This generalization also allows \tool to
  integrate arithmetic solvers with the
  \egraph---\mapi expressions returned by
  solvers are unified with the \eclass that
  matched the \textbf{\F{List Solve}}
  rule's list pattern.

\begin{figure}
\begin{lstlisting}[
  language=Python,
  basicstyle=\footnotesize\ttfamily,
  xleftmargin=2em,
  literate={.apply}{.apply}{6}]
def Szalinski(csg : core_caddy):
    egraph, root = make_egraph(csg)
    while egraph.changed()
        for (lhs, rhs) in SZALINSKI_REWRITES:
            matches = egraph.search(lhs)
            for (eclass, subst) in matches:
                c = egraph.add(apply(rhs, subst))
                egraph.unify(eclass, c)
    return egraph.extract(root, min_size)
\end{lstlisting}
  \caption{
    Equality Saturation for \caddy in \tool
%%    \tool implements Equality Saturation to
%%      shrink \caddy programs using rewrites
%%      that expose repetition in
%%      noisy Phase 1 decompiler outputs
%%      and reroll loops.
  }
  \label{fig:alg}
\end{figure}

%% \tool's rewriting algorithm is inspired by equality saturation~\cite{eqsat}.
%% Equality saturation is a technique that uses
%%   \egraphs and rewrite rules for program optimization.
%% \autoref{fig:alg} shows its pseudocode.

\subsection{Equality Saturation in \tool}
%\todo{at its core, \tool uses an \egraph to implement eqsat}
\tool implements Equality Saturation~\cite{eqsat}
  for \caddy (\autoref{fig:alg}).
First, an \egraph is created from the input
  Core \caddy expression.
%%  initial expression by
%%  placing every \enode in its own \eclass
%%  and recording the \eclass that represents that expression.
Then \tool expands the \egraph by
  repeatedly applying rewrites.
Searching the \egraph for a
  rewrite's left-hand side pattern results in
  a list of (\eclass, substitution) pairs that
  indicate where and how a pattern was matched.
For each pair $(c, \phi)$,
  \tool generates an expression $e$ by
    applying the rewrite's right-hand side function to $\phi$,
  adding $e$ to the \egraph yielding \eclass $c'$, and
  unifying $c$ and $c'$.
\tool continues applying rewrites until
  the \egraph \textit{saturates}
  (reaches a fixpoint where no rewrites further expand the \egraph),
  or a timeout is reached.
In the case of saturation,
  \tool has discovered \textit{all} equivalences
  derivable from its rewrites.

%\todo{\tool continues applying rewrites until they no longer expans the eg.
%this fixed point is known as saturation. when an eg is saturated, the algorithm
%has discovred all equivalences discoverable from the rws. In some cases,
%saturation takes too long so we impose a timeout.}

Finally, \tool extracts the smallest
  \caddy program represented by the
  initial Core \caddy input's \eclass in a simple
  bottom-up traversal of the \egraph.
\tool uses program size as a proxy for editability.
%  which we have qualitatively been satisfied in our own use of the tool.
Past work provides extraction strategies for
  various kinds of cost functions~\cite{eqsat, herbie},
  but we leave further exploration of CAD cost functions
  in \tool to future work.

%% The \textit{extraction} procedure then chooses a represented
%%   expression from the \egraph that minimizes a given cost function.
%% The complexity of extraction varies with the structure of the cost
%%   function.
%% \tool uses a simple monotonic cost function (weighted AST size),
%%   so a greedy
%%   extraction algorithm similar to that in other work \cite{herbie} using \egraphs
%%   suffices to extract the minimum cost represented expression.
%%
%% Since \tool uses a generalized equality saturation strategy
%% for implementing its rewrite system,
%% all the rules in \autoref{fig:loop-rules}, \autoref{fig:cad-rules}
%% are applied in every iteration \todo{ until the \egraph saturates, need to explain
%% that this is modulo heuristics}.

\subsection{Structure Finding in \egraphs}
\label{subsec:struct-eg}

Since \tool's rewrites contain CAD identities that
can fire in every iteration,
the structure finding procedure as
presented in \autoref{subsec:struct} must be
  enhanced.
It must consider that multiple affine transformations
  may be introduced in the same \eclass by the CAD identities.
Given a list of \eclasses $e_1,e_2,...,e_n$, the structure finder aims to
  extract \F{Map2s} that remove one level of structure.
However, due to rules like \F{Affine Combination} from
  \autoref{fig:cad-rules}, each eclass may contain multiple equivalent
  enodes with the same affine operation.
If \eclass $e_i$ has 2 \enodes with the \rotate operator, for example,
  the structure finder can choose from 2 different \F{Rotates} at
  each of the $n$ \eclasses in the list.
Each of these $2^n$ \F{Map2s} has
  distinct children, and will therefore be a distinct \enode in the
  \egraph, all unified in the same \eclass as the list itself.
\tool must operate on large lists of Core \caddy programs,
  but such an exponential number of \enodes would blow up the \egraph.

%\tool instead capitalizes on the observation that equivalences of
%  similar affine transformations will have a similar structure.
\tool instead capitalizes on the observation that it is not useful
  to pick different affine enodes within similar-looking eclasses.
Consider again the ship's wheel example presented in
  \autoref{subsec:eg-motiv}.
After applying the two \rotate identities from \autoref{fig:cad-rules},
  the \eclasses for the top-level affines in the list contain the following
  enodes (one \eclass per row, \enodes shown with their parameters for clarity):
\begin{center}
\begin{tabular}{l@{\hspace{2mm}}l@{\hspace{4mm}}c@{\hspace{4mm}}c}
%\begin{lstlisting}[basicstyle=\scriptsize\sffamily]
% \setlength{\tabcolsep}{40pt}
\textbf{\F{$a$:}} & \F{(\trans~[1,-0.5,0] $x_1$)} & \F{(\rotate~[0,0,0] $a$)} & \\[-1mm]
\textbf{\F{$b$:}} & \F{(\rotate~[0,0,60] $x_2$)} & \F{(\rotate~[0,0,0] $b$)} & \\[-1mm]
\textbf{\F{$c$:}} & \F{(\rotate~[0,0,120] $x_3$)}  & \F{(\rotate~[0,0,0] $c$)} & \\[-1mm]
\textbf{\F{$d$:}} & \F{(\scale~[-1,-1,1] $x_4$)}  & \F{(\rotate~[0,0,0] $d$)} & \F{(\rotate~[0,0,180] $x_4$)} \\[-1mm]
\textbf{\F{$e$:}} & \F{(\rotate~[0,0,240] $x_5$)} & \F{(\rotate~[0,0,0] $e$)} & \\[-1mm]
\textbf{\F{$f$:}} & \F{(\rotate~[0,0,300] $x_6$)} & \F{(\rotate~[0,0,0] $f$)} &
\end{tabular}
\end{center}
%\end{lstlisting}

The structure finder calculates the \emph{affine signature} of each
  \eclass as the multiset of the kinds affine operators in the \eclass.
In the above example,
  \eclass \textbf{\F{$a$}}'s affine signature is $\{\trans, \rotate\}$,
  \textbf{\F{$d$}}'s is $\{\scale, \rotate, \rotate\}$,
  and the others all share the same signature: $\{\rotate, \rotate\}$.
  % the four \eclasses that have two \rotate operators have the same affine signature: \{(\rotate, 2)\},
  % and the \eclass with the \scale has affine signature: \{(\scale, 1), (\rotate, 2)\}.
A \textit{group} is a set of \eclasses that
  share the same affine signature.
When trying to extract a \rotate, the structure finder
  will \emph{not} take the Cartesian product of the \F{Rotates} in each
  \eclass---doing so would lead to $2^{5}$ possible ways to combine \rotate.
%Instead, it groups the \eclasses by affine signature, and
%  takes the \emph{same} affine from each group
%  (using their order in the \eclass).
Instead, it takes the Cartesian product of affine choices for each
  group, and extends the \textit{same} choice of affine over
  all \eclasses within the group
  (using the order of affines in the \eclasses).
In this example, the only affine that can be extracted is \F{\rotate},
  since the other affines do not appear in the affine signature of all groups.
For the \rotate affine,
  group \textbf{\F{$a$}} has one choice,
  group \textbf{\F{$d$}} has 2 choices, and
  group \textbf{\F{$ b, c, e, f $}} also has 2 choices.
%For the \eclasses with the same affine signature,
%the structure finder with either take the first \rotate for all of them or the second.
%The first \eclass only has one \rotate,
%  and the fourth has two, resulting in two additional choices.
This reduces the number of \F{(Map2 Rotate ...)}
  expressions introduced from
  $2^{5} = 32$ to 4.

\section{\Semtag}
\label{sec:inverse}

\egraphs and CAD rewrites allow \tool to
  expose repetitive structure and reroll loops
  even when a Core \caddy input
  exhibits obfuscating variations
  (e.g., \F{Scale [-1,-1,1]} instead of \F{Rotate [0,0,180]}).
However, existing mesh decompilers tend to also
  order and group CAD subexpressions
  by geometric proximity or other heuristics that,
  from \tool's perspective,
  make recovering high-level structure challenging.
Unless the right reordering and regrouping of subexpressions
  can be found, list solvers will
  fail to infer \mapis and \tool
  will be unable to reroll loops and shrink \caddy programs.

To address this challenge,
  we introduce \emph{\semtag},
  a novel way for solvers to optimistically
  unify expressions in an \egraph that
  would be equivalent modulo reordering or regrouping.

%% \egraphs allow \tool's structure finding and list solving to work even
%%   when repetitive structure is hidden behind CAD identities.
%% However, these techniques alone cannot handle outputs of existing CAD
%%   decompilers that arbitrarily\footnotemark{} reorder and regroup
%%   subexpressions.
%% \footnotetext{
%%   \mtoc tools order/group subexpressions by geometric proximity or other
%%   heuristics that are arbitrary from the perspective of Stage 2 tools
%%   like \tool.
%% }
%% \tool requires
%%   a novel technique called \emph{\semtag} handle to inputs with these
%%   troublesome characteristics.

% Consider finally the representative \mtoc output for the ship's wheel
%   in \autoref{fig:ship-csg-bad}.
% Using the techniques discussed in Sections \ref{sec:loops} and
%   \ref{sec:egraphs}, \tool will find the following represented in the
%   \egraph:

\begin{figure}
\begin{lstlisting}[xleftmargin=2mm,style=rewrite]
    (Fold Union (List
      (Rotate |$[0,0,120]$| (Translate |$[1,-0.5,0]$| (Cuboid |$[10,1,1]$|)))
      (Rotate |$[0,0,0]$|   (Translate |$[1,-0.5,0]$| (Cuboid |$[10,1,1]$|)))
      (Rotate |$[0,0,300]$| (Translate |$[1,-0.5,0]$| (Cuboid |$[10,1,1]$|)))
      (Cylinder |$[1,5]$|)
      (Rotate |$[0,0,180]$| (Translate |$[1,-0.5,0]$| (Cuboid |$[10,1,1]$|)))
      (Rotate |$[0,0,240]$| (Translate |$[1,-0.5,0]$| (Cuboid |$[10,1,1]$|)))
      (Rotate |$[0,0,60]$|  (Translate |$[1,-0.5,0]$| (Cuboid |$[10,1,1]$|)))))
\end{lstlisting}
\caption{
  Section~\ref{sec:loops}~and~\ref{sec:egraphs} techniques
  find the ``\rotate then \trans'' structure
  from the realistic \autoref{fig:ship-csg-bad}.
  Without \semtag, loop rerolling now gets stuck.
}
\label{fig:almost-there}
\end{figure}

\autoref{fig:almost-there} shows
  how far CAD rewrites combined with techniques from
  previous sections get for the \autoref{fig:ship-csg-bad} example.
Unfortunately, \cyl is still
  \F{Union}ed with \F{Cuboids},
  preventing the structure finder from
  pulling out the \rotate.
Even if the \cyl were removed,
  the list order would prevent
  solvers from inferring a \mapi for
  the \F{Rotate} parameters.

%% Once again considering the ship's wheel model (\autoref{fig:almost-there}),
%% the CAD rewrites in the \egraph are able to expose the similar
%%   \rotate then \trans manipulation on the \F{Cuboids}.
%% Unfortunately, the \cyl is still \F{Union}ed together with the
%%   \F{Cuboids}, preventing the structure finder from pulling out the
%%   \rotate.
%% Even if the \cyl were removed,
%%   the arithmetic solver could not solve for the parameters of the
%%   \F{Rotates} since they are in the wrong order.

%% Of course, if they correct regrouping and reordering was somehow in
%%   the \egraph, then the structure finder and arithmetic would work as
%%   expected.

Unlike the previous section,
  adding more rewrites does not help.\footnote{
    We can report that AC-matching is
    a problem both in theory and practice.}
  %(i.e., commutativity and associativity)
\egraphs do \emph{not} compactly represent equivalences
  due to reordering associative and commutative operators
  like \union. \shepherd{This is known as the AC-matching problem~\cite{acmatch1}
  (A stands for associativity, and C for commutativity)
  and it prevents efficiently exploring all
  possible reorderings and regroupings.}
%\todo{do we need to cite acmatch2 separately? mentioning canonicalization doens't seem helpful}

%% Unlike the case in the previous section with CAD identities, simply
%%   adding rewrites like commutativity and associativity will not suffice.
%% \egraphs do \emph{not} compactly encode these identities; they result in an a
%%   exponential number of \eclasses, one for each permutation of
%%   arguments to an associative commutative operator like \union.
%% This AC-matching \cite{acmatch1} effectively rules out considering all
%%   possible reorderings and regroupings.
%% \todo{do we need to cite acmatch2 separately? mentioning canonicalization doens't seem helpful}

\tool addresses this with a new technique, \textit{\semtag},
  that allows solvers to
  speculatively transform their inputs to allow for more
  profitable rewriting.
A solver that cannot simplify input $A$ may,
  for some transformation $F$,
  be able simplify $F(A)$ to $B$.
\Semtag simply allows the solver to
  ``wrap'' $B$ with $F^{-1}$ before unifying it with $A$,
  even though $A$ and $B$ are not equivalent.

\Semtag enable locally-reasoning solvers to
  register potentially profitable
  regroupings and reorderings in an \egraph.
Simple syntactic rewrites then
  propagate these ``hints'' globally through the \egraph,
  allowing other solvers to try them,
  and contextually eliminate inverse transformations when possible
  (e.g., under order-insensitive operations like \F{Fold Union}).

\subsection{Extended \caddy}
\label{subsec:ecaddy}

Extended \caddy (\autoref{fig:grammar2}~and~\ref{fig:sem2})
  adds \semtag that allow solvers to
  record how they manipulated their input.
These extended forms are only introduced in the \egraph;
  \tool's cost function ensures extraction produces
  regular \caddy programs.
Semantically, these constructs either
  undo the transformation performed by the solver to recover the input, or
  perform the transformation on some other part of the program.
\ssort and \unsort take a permutation $p$ and a list $\ell$,
  imposing (respectively, undoing) $p$ on $\ell$.
\spart takes a partitioning $P$ (a list of lengths) and a list $\ell$,
  breaking down $\ell$ into a list of sublists according to $P$.
\unpart takes a partitioning and a list of lists and flattens the
  latter; the partitioning is only use to propagate information.
\polar and \unpolar take a 3D vector $c$ and a list of 3D vectors in spherical
  coordinates about $c$, returning a list of the vectors in Cartesian
  coordinates (and vice versa).

% We present Extended \caddy, which provides \semtag in addition to the
% features of \caddy.
% shows the syntax of Extended \caddy---
% it supports five \semtag, \ssort, \unsort, \spart, \unpart, and \unpolar.
% The first four are used for list manipulation, in particular, sorting
% and partitioning.
% Intuitively, \ssort/\unsort and \spart/\unpart perform the
% opposite manipulation.
% \ssort sorts a list according to a
% \emph{\F{permutation}}, which is a set of indices that indicate how a list
% should be sorted. \spart partitions a list
% into sublists of lengths that are encoded in \emph{\F{partitioning}}.
% \autoref{fig:sem2} shows the semantics of \ssort/\unsort and \spart/\unpart.
% In the subsequent sections, we discuss some representative rewrite
% rules that demonstrate the strength of the \semtag technique.

\begin{figure}
  \sf
  \begin{grammar}
    <permutation> ::= $\langle$ n, n, ... $\rangle$ \quad
    partitioning ::= $\langle$ n, n, ... $\rangle$

    <inv> ::= (Sort <permutation> <*-list>)
    \alt (Unsort <permutation> <*-list>)
    \alt (Part <partitioning> <*-list>)
    \alt (Unpart <partitioning> <*-list>)
    \alt (Spherical <vec3> <vec3-list>)
    \alt (Unspherical <vec3> <vec3-list>)
  \end{grammar}
 \caption{ Syntax of Extended \caddy. }
 \label{fig:grammar2}
\end{figure}

\begin{figure}
  \centering
  \footnotesize
  \newcommand\itrm[1]{\rm \it #1}

  \begin{equation*}
  \begin{array}{c}
  \inference[]
    {e \Rightarrow \F{(List $v_1$ $v_2$ ... $v_n$)}}
    {\F{(Sort $\langle i_1, i_2, ..., i_n \rangle$ e)} \Rightarrow \F{(List $v_{i_1}$ $v_{i_2}$ ... $v_{i_n}$)}}
    \vspace{2em} \\

  \inference[]
    {e \Rightarrow \F{(List $v_{i_1}$ $v_{i_2}$ ... $v_{i_n}$)}}
    {\F{(Unsort $\langle i_1, i_2, ..., i_n\rangle$ e)} \Rightarrow \F{(List $v_1$ $v_2$ ... $v_n$)}}
    \vspace{2em}\\

  \inference[]
    {
     \mathit{sum_0} = 0 \hspace{49.5pt}
     \mathit{sublist}_i = \F{(List $v_{\mathit{sum}_{i-1}}$ ... $v_{\mathit{sum}_i}$)} \\
     \mathit{sum}_i = \mathit{sum}_{i-1} + l_i \hspace{27pt}
     e \Rightarrow \F{(List $v_1$ $v_2$ ... $v_{\mathit{sum}_n}$)} \hspace{10pt}
    }
    {\F{(Part $\langle l_1, l_2, ..., l_n\rangle$ e)}
     \Rightarrow
     \F{(List $\mathit{sublist}_1$ ... $\mathit{sublist}_n$)}}

    \vspace{2em}\\

    \inference[]
    {
     \mathit{sum_0} = 0 \hspace{44.5pt}
     \mathit{sublist}_i = \F{(List $v_{\mathit{sum}_{i-1}}$ ... $v_{\mathit{sum}_i}$)} \\
     \mathit{sum}_i = \mathit{sum}_{i-1} + l_i \hspace{20pt}
     e \Rightarrow
      \F{(List $\mathit{sublist}_1$ ... $\mathit{sublist}_n$)}
    }
    {
    \F{(Unpart $\langle l_1, l_2, ..., l_n\rangle$ e)}
    \Rightarrow
    \F{(List $v_1$ $v_2$ ... $v_{\mathit{sum}_n}$)}
    }

    \vspace{2em}\\
    \inference[]
    {
      e \Rightarrow \F{(List $v'_1$ $v'_2$ ... $v'_n$)} \qquad
      \F{$v_i$ = to_spherical ($\mathit{center}$, $v'_i$)}
    }
    {
      \F{(Spherical n $\mathit{center}$ e)}
    \Rightarrow
    \F{(List $v_1$ $v_2$ ... $v_n$)}
    }

    \vspace{2em}\\
    \inference[]
    {
      e \Rightarrow \F{(List $v'_1$ $v'_2$ ... $v'_n$)} \qquad
      \F{$v_i$ = to_cartesian ($\mathit{center}$, $v'_i$)}
    }
    {
      \F{(Unspherical n $\mathit{center}$ e)}
    \Rightarrow
    \F{(List $v_1$ $v_2$ ... $v_n$)}
    }

  \end{array}
  \end{equation*}
  \caption{
    Big step semantics for Extended \caddy.
  }
  \label{fig:sem2}
\end{figure}

%%% Local Variables:
%%% TeX-master: "main"
%%% End:

\subsection{Restructuring with \unpart and \unsort}
\label{subsec:unsort}
\label{subsec:part}

Using \semtag, \tool can finally get the desired output given the
  realistic input for the ship's wheel (\autoref{fig:ship-csg-bad}).
Starting from \autoref{fig:almost-there},
  \tool separates the \cyl from the \F{Cuboids} with partitioning and
  sorts the list of \F{Cuboids} on their \rotate parameters,
  revealing repetitive structure similar to
  the ideal input (\autoref{fig:ship-csg-good}).

\paragraph{Partitioning}
\tool includes a \emph{partitioning solver} that
  uses \semtag and a set of heuristics to
  restructure lists in ways that group
  similar list elements together
  (e.g., by kind of geometric primitive).
The partitioner can split up elements of a list by
  equivalence class,
  individual components of 3D vectors,
  and kinds of affine transformations.
In \autoref{fig:almost-there}, the partitioner will
  split the list into:
\begin{lstlisting}[xleftmargin=0mm,basicstyle=\sffamily\scriptsize]
(Fold Union
  (Unpart $\langle 1, 6 \rangle$
    (List (Cylinder |$[1,5]$|))
    (List (Rotate |$[0,0,120]$| (Translate |$[1,-0.5,0]$| (Cuboid |$[10,1,1]$|)))
          (Rotate |$[0,0,0]$|   (Translate |$[1,-0.5,0]$| (Cuboid |$[10,1,1]$|)))
          (Rotate |$[0,0,300]$| (Translate |$[1,-0.5,0]$| (Cuboid |$[10,1,1]$|)))
          (Rotate |$[0,0,180]$| (Translate |$[1,-0.5,0]$| (Cuboid |$[10,1,1]$|)))
          (Rotate |$[0,0,240]$| (Translate |$[1,-0.5,0]$| (Cuboid |$[10,1,1]$|)))
          (Rotate |$[0,0,60]$|  (Translate |$[1,-0.5,0]$| (Cuboid |$[10,1,1]$|)))))
\end{lstlisting}

The introduced \unpart is equivalent to \concat,
  but additionally stores partitioning hints.
Now that the \F{Rotates} are gathered uniformly in a list,
  the structure finder will rewrite the list to:

\noindent
\begin{lstlisting}[xleftmargin=1.5mm,basicstyle=\sffamily\scriptsize]
(Map2 Rotate
    (List |$[0,0,120]$| |$[0,0,0]$| |$[0,0,300]$| |$[0,0,180]$| |$[0,0,240]$| |$[0,0,60]$|)
    (Repeat 6 (Translate |$[1,-0.5,0]$| (Cuboid |$[10,1,1]$|))))
\end{lstlisting}

The arithmetic solver from \autoref{subsec:solve}
  cannot find a closed form for
  this list of \rotate parameters.
The solver could, however,
  find a closed form if it were free to sort the list
  (by $z$-coordinate, in this case).
The sorted list is \textbf{not} equivalent to the original.
Since the solver only rewrites locally,
  it does not know if the list appears under
  a \F{Fold Union} (which is AC) or
  a \F{Fold Diff} (which is \emph{not} AC).
In the \egraph,
  \textit{both} situations could actually
  hold due to sharing.
The solver \emph{cannot} soundly rewrite the
  original list to the closed form \mapi,
  but it \emph{can} soundly rewrite the list to:

\noindent
\begin{lstlisting}[xleftmargin=1.5mm,basicstyle=\sffamily\scriptsize]
  (Unsort $\langle 1, 5, 0, 3, 4, 2\rangle$ (Tabulate (i 6) [0, 0, 60i]))
\end{lstlisting}

The \unsort \semt allows the solver to introduce
  the closed form \mapi in the \egraph,
  but \tool will never extract it or any other program
  using the \semt forms from Extended Caddy.
Instead,
  rewrites propagate \semts between
  invocations of locally-reasoning solvers,
  and additional rules eliminate \semtag in
  contexts invariant to the relevant transformation;
  these rules are shown in \autoref{fig:inv-rules}.
The \textbf{\F{Map2 Unsort Params}} rewrite applies
  to our running example, producing:
\noindent
\begin{lstlisting}[xleftmargin=1.5mm,basicstyle=\sffamily\scriptsize]
(Unsort $\langle 1, 5, 0, 3, 4, 2 \rangle$ (Sort $\langle 1, 5, 0, 3, 4, 2 \rangle$
  (Map2 Rotate
    (Unsort $\langle 1, 5, 0, 3, 4, 2\rangle$ (Tabulate (i 6) [0, 0, 60i]))
    (Repeat 6 (Translate |$[1,-0.5,0]$| (Cuboid |$[10,1,1]$|))))))
\end{lstlisting}

Semantically, this is no different, as \F{(Unsort $p$ (Sort $p$ $x$)) = $x$},
  but since the \map is in the same \eclass as the original list of
  \F{Rotates}, the \F{\bf Sort Application} rule can fire,
  communicating the profitable ordering of the \rotate parameters to the outer list.
Now, the structure finder and arithmetic solver apply to the sorted
  list of \F{Rotates}, bringing the whole program to:
\begin{lstlisting}[xleftmargin=5mm,basicstyle=\sffamily\scriptsize]
(Fold Union
  (Unpart $\langle 1,6 \rangle$
    (List (Cylinder |$[1,5]$|))
    (Unsort $\langle 1, 5, 0, 3, 4, 2 \rangle$
      (Tabulate (i 6)
        (Rotate [0, 0, 60i]
          (Translate |$[1,-0.5,0]$|
            (Cuboid |$[10,1,1]$|))))))))
\end{lstlisting}

From here an additional rewrite (elided from \autoref{fig:inv-rules})
  can lift the \unsort over the \unpart:
\begin{lstlisting}[xleftmargin=5mm,basicstyle=\sffamily\scriptsize]
(Fold Union
  (Unsort $\langle 0, 2, 6, 1, 4, 5, 3 \rangle$
    (Unpart $\langle 1, 6 \rangle$
      (List (Cylinder |$[1,5]$|))
      (Tabulate (i 6)
        (Rotate [0, 0, 60i]
          (Translate |$[1,-0.5,0]$|
            (Cuboid |$[10,1,1]$|))))))))
\end{lstlisting}

Next, the \F{\bf Unsort Elimination} rule removes the \unsort,
  since \F{Fold Union} is invariant to order.
Finally one additional rule that transforms a
  \union of an \unpart into a \union of \unions (not shown),
  produces the desired \caddy output (\autoref{fig:ship-prog}).

\begin{figure}
\begin{tabular}{lll}
\rewritenamecond{Map2 Unsort Params}{- \textit{cads} rule analogous} \\
    \begin{lstlisting}[style=rewrite]
    (Map2 $\mathit{affine}$
      (Unsort $\mathit{perm}$ $\mathit{params}$)
        $\mathit{cads}$)
    \end{lstlisting}
    & \rewritesto &
    \begin{lstlisting}[style=rewrite]
    (Unsort $\mathit{perm}$ (Sort $\mathit{perm}$
      (Map2 $\mathit{affine}$
        (Unsort $\mathit{perm}$ $\mathit{params}$)
          $\mathit{cads}$)))
    \end{lstlisting}
\rewritespace
\\
\rewritename{Sort Application} \\
    \begin{lstlisting}[style=rewrite]
    (Sort $\langle i_1, ... i_n \rangle$ (List $x_1$ ... $x_n$))
    \end{lstlisting}
    & \rewritesto &
    \begin{lstlisting}[style=rewrite]
    (List $x_{i_1}$ ... $x_{i_n}$)
    \end{lstlisting}
\rewritespace
\\
\rewritename{Unsort Elimination} \\
    \begin{lstlisting}[style=rewrite]
    (Fold Union (Unsort $\mathit{perm}$ $l$))
    \end{lstlisting}
    & \rewritesto &
    \begin{lstlisting}[style=rewrite]
    (Fold Union $l$)
    \end{lstlisting}
    \\
    \begin{lstlisting}[style=rewrite]
    (Unsort $\mathit{perm}$ (Repeat $n$ $x$))
    \end{lstlisting}
    & \rewritesto &
    \begin{lstlisting}[style=rewrite]
    (Repeat $n$ $x$)
    \end{lstlisting}
\rewritespace
\\
\rewritenamecond{Map2 Unpart Cads}{- \textit{params} rule analogous} \\
    \begin{lstlisting}[style=rewrite]
    (Map2 $\mathit{affine}$
      $\mathit{params}$
      (Unpart $\mathit{part}$ $\mathit{cads}$))
    \end{lstlisting}
    & \rewritesto &
    \begin{lstlisting}[style=rewrite]
    (Unpart $\mathit{part}$ (Part $\mathit{part}$
      (Map2 $\mathit{affine}$
        $\mathit{params}$
        (Unpart $\mathit{part}$ $\mathit{cads}$))))
    \end{lstlisting}
\rewritespace
\\
\rewritename{Unpart to Concat} \\
    \begin{lstlisting}[style=rewrite]
    (Unpart $\mathit{part}$ $\mathit{lists}$)
    \end{lstlisting}
    & \rewritesto &
    \begin{lstlisting}[style=rewrite]
    (Concat $\mathit{lists}$)
    \end{lstlisting}
\rewritespace
\\
\rewritename{Unspherical Trans} \\
    \begin{lstlisting}[style=rewrite]
    (Map2 Trans
      (Unspherical $n$ $\mathit{center}$ $\mathit{params}$)
      $\mathit{cads}$)
    \end{lstlisting}
    & \rewritesto &
    \begin{lstlisting}[style=rewrite]
    (Map2 Trans
      (Repeat $n$ $\mathit{center}$)
      (Map2 TranslateSpherical $\mathit{params}$ $\mathit{cads}$))
    \end{lstlisting}
\rewritespace
\\
\end{tabular}
\caption{Representative set of rewrite rules for propagation and elimination of \semtag.}
\label{fig:inv-rules}
\end{figure}

\subsection{Solving for Spherical Coordinates}
\label{subsec:unpolar}
%Consider the grid of spheres in \autoref{fig:flower}.
%Following are the vectors representing one of the clusters: \\
%\F{[(10, 10, 0), (10, 5, 0), (5, 10, 0), (5, 5, 0)]}.
%While a doubly-nested
%loop is one potential parametrization for this list, it is most useful
%for adding or removing a row or column. It is not useful
%for example, to convert it to pentagonal grid of spheres.
%\begin{figure}[h]
%    \includegraphics[width=0.5\linewidth]{fig/flower.png}
%  \caption{4 groups of 4 spheres.}
%  \label{fig:flower}
%\end{figure}
%Instead, a better
%solution is a function that represents the vectors in the
%spherical coordinate system.
\Semtag are not restricted to list manipulations.
In addition to sorting,
  \tool's arithmetic solvers can
  convert lists to spherical coordinates~\cite{spherical}.
The resulting list may be easier to find a closed form \mapi for,
  but it is not equivalent to the input.
% takes the list of 3D vectors and converts them to spherical coordinates
% The polarizer
% returns a list of vectors in spherical coordinates. It also returns the
% \textit{center} used for the computation (our implementation uses centroid).
% This new list of spherical vectors is then used by the arithmetic
% solver to find a closed form which may be a better function for certain edits to a CAD design.
% However, similar
% to list permutations, the expression returned by the solver
% will not be equivalent to the original list.
Therefore, the solver wraps the \mapi
  in an \semt, \unpolar, before passing it to
  the \egraph for unification.
If the \unpolar propagates under a \trans, then
  the \F{Unspherical Trans} rule can replace it with \F{TranslateSpherical}
  form.
This approach allows \tool to solve for closed forms of lists in
  spherical coordinates without the solver knowing whether or not it
  is solving for a list of \trans parameters.
% \unpolar is then eliminated in the \egraph using
% in \autoref{fig:inv-rules}.
%  Intuitively, \unpolar communicates to the \egraph,
% how to get back the original concrete list from the solution it found---use the
% center to undo the computations in \F{polarize} for all the list elements.
% The semantics of \unpolar are in \autoref{fig:sem2}.
% The expression it returns is of the form:
% \F{(Unpolar n center (Tabulate (i n) f))}
% \todo{show actual f},
% where \F{n} is the number of elements in the list.
%In this function, changing
%\todo{show some change} changes the shape from a square grid to a pentagonal
%grid, which is much harder and unintuitive in a nested loop based function.
%The \semtag,

%Recall that \unpolar is generated when
%\tool's solver finds a closed form for the spherical representation of
%Cartesian vectors. \F{unpolar_trans} in \autoref{fig:rws} shows how
%\unpolar is eliminated
%by first rewriting \map's list of \caddy expressions
%using \polar with the arguments returned by the solver (\F{params}),
%and then un-polarizing this new list by an outer \trans which
%translates all its elements by \F{center} to convert the vectors back to
%Cartesian coordinates.
%This is the only
%propagation rule required for
%\unpolar because it only applies to \trans.

\subsection{Inverse Transformations, Broadly}
\label{subsec:semtagbroad}

This section and our evaluation show that \semtag are
  effective for shrinking \caddy programs, but the technique could
  be applied more broadly to other uses of Equality Saturation.
The key insight is that solvers can remain simple because they only
  have to reason locally.
They are given the flexibility to speculate on potentially profitable
  ways to transform their inputs.
Rewrites can then propagate this information and contextually
  eliminate the transformations.
As in traditional Equality Saturation, these rewrites (and now simple
  solvers) compose in emergent ways, leading to unexpectedly powerful
  outcomes, that would have otherwise required more complicated
  solvers with deep, contextual reasoning ability.

%%% Local Variables:
%%% TeX-master: "main"
%%% End:

\section{Implementation}
\label{sec:impl}

\tool is implemented in 3000 lines of \rust
  and uses \texttt{egg}~\cite{egg}, an open source \egraph library.
%The implementation consists of defining \cadilac together with a cost function
%which is used for extracting a simplified program from the \egraph,
%implementing the rule database including \semts, implementing the arithmetic solver
%and partitioner, and an evaluator (described in the next section).
\autoref{tab:impl} provides a break down of
the LOC for each of \tool's components.
\tool uses only simple, custom solvers for arithmetic and list
partitioning.
The most of \tool's 65 rewrites are syntactic and compactly expressed,
and the remainder either call out to the solvers or manipulate lists.
\tool is publicly available at
\textit{https://github.com/uwplse/szalinski.git}.

\begin{table}[h]
  \begin{tabular}{ccccc}
    % \toprule
    \cadilac & Rewrites & Solvers & Main loop & Validation \\
    \hline
    300 & 900 & 400 & 300 & 1100
    % \bottomrule
  \end{tabular}
  \caption{Approximate LOC breakdown of \tool}
  \label{tab:impl}
\end{table}

% \autoref{fig:sz-tool} presents the workflow of \tool.
% It takes a flat CSG program input, and converts it into Core \caddy.
% Most of the inputs are in OpenSCAD~\cite{openscad}'s CSG syntax,
% which is different from Core \caddy's syntax. For example, OpenSCAD has
% cubes that are parametrized by different side lengths to obtain cuboids,
% and has cylinders that are parametrized by the number of sides
% to get shapes like hexagonal prisms.
% \tool's parser parses the inputs to Core \caddy.
% which initializes the \egraph.
% \tool iteratively applies rewrites
% and uses a cost function (AST size) to extract
% the smallest program.
% \todo{talk about nested loops? or solvers implementation details?}

\subsection*{Correctness}
\label{subsec:correct}

\begin{figure}
  \centering
  \includegraphics[width=\linewidth]{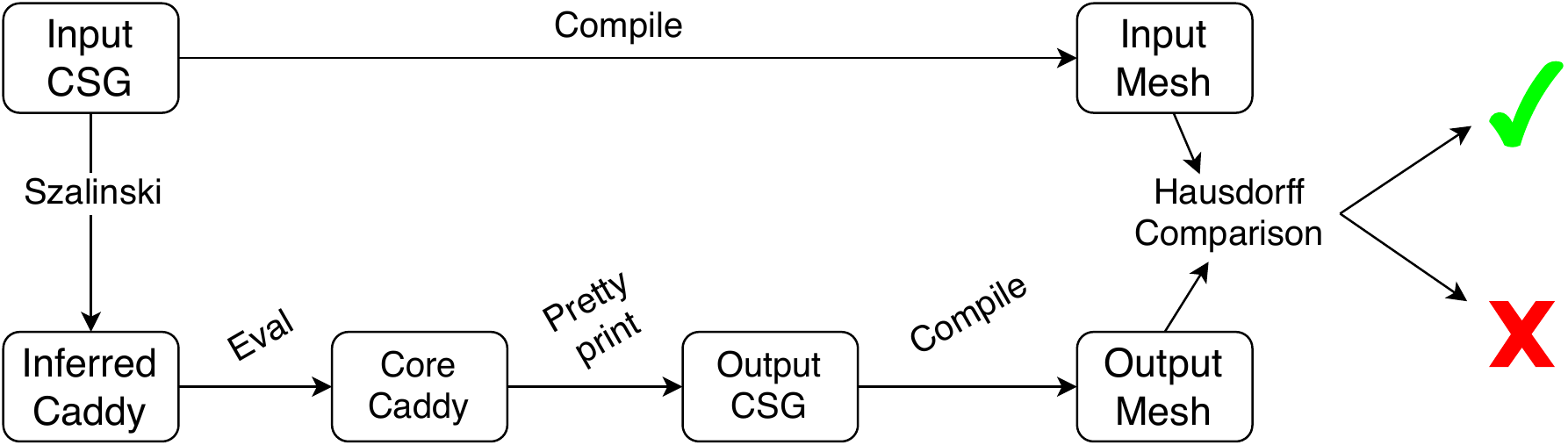}
  \caption{
    The \tool tool.
    The simplification process
    outputs a parameterized program in the \caddy language.
    The validation step evaluates \tool's output to Core \caddy, pretty prints
    it to CSG and uses an open source CAD compiler to generate a mesh. The
    input to \tool is also compiled to a mesh. The two meshes are then compared
    using Hausdorff distance.
  }
  \label{fig:sz-tool}
\end{figure}

To validate \tool's correctness, we test that the initial and final
\caddy programs compile to similar meshes (\autoref{fig:sz-tool}).
% We implemented a verifier
% that ensures that \tool's output is correct
\tool first evaluates a \caddy program back to a flat Core \caddy
  program which is then pretty printed to a CSG program.
We use the open source OpenSCAD~\cite{openscad} tool to compile
  the CSGs to triangular meshes.
We then use the CGAL~\cite{cgal} library to compute the Hausdorff
  distance~\cite{hausdorff, inverse} between the two meshes.
A Hausdorff distance less than a small $\epsilon$ indicates
  equivalence (ideally it should be zero, but due to rounding errors,
  it is sufficient to check against $\epsilon$).
% In our experiments, $\epsilon$ was set to 0.001.
% Due to numerical inaccuracies and timeouts,
% we found that CGAL's Hausdorff computation failed for 148 examples.
% For these examples, we manually
% inspected the input and the output mesh to judge their equivalence.

%% \todo{ztatlock this is orphan from section 2}
%% \todo{ztatlock: move the content below to eval, maybe cut later}
%% \todo{from zach: For \tool, the behavior of a Core \caddy program is

%%   the mesh it compiles it to.\footnote{To evaluate \tool we used the OpenScad compiler.}
%% We consider Caddy programs equivalent
%%   if they evaluate and then compile to ``similar'' meshes,
%%   even if they do not evaluate to the same Core \caddy program.
%% We must settle for ``similar'' meshes, as floating point imprecision
%%   inside CAD compilers renders CSGs that are equivalent according to
%%   geometric semantics \cite{reincarnate} into different meshes.
%% Our evaluation (\autoref{sec:eval}) describes how we measure mesh
%%   similarity.
%% }

%%% Local Variables:
%%% TeX-master: "main"
%%% End:

%!TEX root = main.tex
\section{Evaluation}
\label{sec:eval}

In evaluating \tool, were interested in the following research questions:

\begin{itemize}

\item \textit{End-to-End.} (\autoref{subsec:end2end})
Does \tool compose with prior mesh decompilation
    tools and find parametrizable programs from
the flat CSG expressions generated by the latter?

\item \textit{Scalability.} (\autoref{subsec:exper})
Does \tool scale to large flat CSGs?
How fast can it find equivalent smaller \caddy programs?

\item \textit{Sensitivity analysis.}
How do the different components of \tool, in particular CAD rewrites
and \semtag, affect its results?

\end{itemize}

\shepherd{We ran our evaluation on a 6 core Intel i7-8700K processor with 32 GB of RAM.}

%\begin{definition}
%
%  Let $compile : \cadilac \rightarrow mesh$ be a function that generates a
%  triangular polygon mesh from a flat CSG program.
%
%\end{definition}
%
%\begin{definition}
%
%  Two meshes, $m_1$, $m_2$ are \textit{Hausdorff equivalent}, i.e. $m_1
%  \equiv_{\mathfrak{h}} m_2$ iff $\delta_\mathfrak{h} (m_1, m_2) \leq \epsilon$ for a small value of
%  $\epsilon$, where $\delta_\mathfrak{h}$ is the Hausdorff distance function.
%
%\end{definition}
%
%\begin{definition}
%
%  Two CSG programs $c$ and $c'$ are equivalent iff
%  $compile (c) \equiv_{\mathfrak{h}} compile(c')$.
%
%\end{definition}
%
%\begin{definition}
%
%  For a flat input CSG program $c$, \tool's parametrized \cadilac output, $c'$ is \textit{correct} if
%  the pretty printed (\autoref{fig:sz-tool}) CSG of $c'$ is equivalent to $c$.
%
%\end{definition}

%\begin{definition}
%  The Hausdorff distance, $\delta_\mathfrak{h}$ between two triangular polygon meshes $m$ and $m'$ is defined by:
%  \begin{align*}
%    \delta_\mathfrak{h}(m, m') = max_{p \in m} (min_{p' \in m'} (\delta(p, p')))
%  \end{align*}
%  where $p$ and $p'$ denote points in $m$ and $m'$ respectively, and $\delta$ is the Euclidean distance function.
%\end{definition}

\subsection{End-to-End Experiments}
\label{subsec:end2end}

To evaluate the composability of \tool with mesh decompilation tools,
  we ran \tool on flat CSGs generated by the
  Reincarnate~\cite{reincarnate} mesh decompiler.
% Our goal was to compare the cost of \tool's \caddy output against the
%   Core \caddy input using \tool's cost function (number of AST nodes).
This required investigating what kinds of models Reincarnate
supports; we found that it worked best on models that do not
contain round edges.
We found 10 such models from Thingiverse~\cite{thing} and ran Reincarnate
on their mesh files to get flat CSGs and converted those to Core \caddy.

Given the Core \caddy inputs, \tool synthesizes \caddy programs
  (\autoref{fig:sz-tool}).
We compared the parametrized programs
synthesized by \tool from Reincarnate's output with manually written
parametrized programs in OpenSCAD (column 1 in \autoref{tab:reincar}).
For four of the 10 models, we found a parametrized
OpenSCAD implementation on Thingiverse.
For the other six, we manually wrote a parametrized implementation in OpenSCAD.
%Since this is an end-to-end evaluation, the parametrized implementation
%in OpenSCAD does not impact the result of \tool in any way.
\autoref{tab:reincar}
shows the comparison of the lines of code  at every stage of the
end-to-end synthesis process,
and the cost of the flat input Core \caddy and the output \caddy.
\tool was able to reduce the cost of the programs by
86\% on average. The last two columns report a sensitivity analysis of \tool on
Reincarnate's output. It shows that both CAD identities and \semtag
contribute significantly to shrinking \caddy programs.

Compiling the \caddy programs to mesh resulted in meshes equivalent to
  the source meshes (Hausdorff distance < 0.001).
We also manually validated that all 10 inferred \caddy programs are
  structurally similar to the parameterized input OpenSCAD programs.

\begin{table}
  \adjustbox{max width=\linewidth} {
  \begin{tabular}{cccccccc}
  \toprule
    %\F{Name} & \F{OpenSCAD LOC} & \F{\# Mesh Triangles} & \F{Core \caddy Cost} & \F{\caddy cost} & \F{No CAD rewrites} & \F{No Inv Trans}\\
    Id & SCAD & \# Tri & $c_{in}$ & $c_{out}$ & No CAD & No Inv\\
  \midrule
     {TackleBox}        & 48 & 280 & 280 &  26 &  60 & 41 \\
     {SDCardRack}       & 13 & 236 & 206 &  26 &  57 & 49 \\
     {SingleRowHolder}  & 10 & 320 & 198 &  16 &  31 & 38 \\
     {CircleCell}       & 14 & 124 & 79  &  16 &  31 & 16 \\
     {CNCBitCase}       & 59 & 268 & 219 &  15 &  27 & 27 \\
     {CassetteStorage}  & 13 & 172 & 141 &  15 &  27 & 25 \\
     {RaspberryPiCover} & 34 & 332 & 271 &  12 &  27 & 32 \\
     {ChargingStation}  & 45 & 192 & 141 &  18 &  27 & 29 \\
     {CardFramer}       & 11 & 200 & 172 &  42 &  83 & 42 \\
     {HexWrenchHolder}  & 13 & 516 & 317 &  16 &  31 & 52 \\
  \midrule
    Average & 26.0 & 264.0 & 202.4 & 20.2 & 40.1 & 35.1 \\
  \bottomrule
  \end{tabular}
 }

\caption{End-to-end evaluation of \tool on the results of
  Reincarnate~\cite{reincarnate}.
  SCAD show LOC in original parametrized OpenSCAD implementations,
  \# Tri shows the number of triangles in the mesh, $c_{in}$ and $c_{out}$
  are the input and output costs.
  The last two columns indicate the cost of the
  output \caddy program when \tool does not apply any CAD identities, and when
  \semtag are turned off, respectively.  }

\label{tab:reincar}
\end{table}

\subsection{Large Scale Evaluation on Thingiverse Models}
\label{subsec:exper}

Mesh decompilation tools have limitations.
Reincarnate for example, works mainly on
shapes without rounded corners and edges.
Therefore, in order to evaluate \tool further,
we performed a larger scale evaluation on models from
Thingiverse~\cite{thing}, a popular online model sharing website.

The goals for this part of the evaluation are: (1) to simulate
the behavior of mesh decompilation tools by flattening
parametrized programs and perturbing them to
reproduce the challenges \chOne and \chTwo (introduced in \autoref{sec:intro}),
and run \tool on these flat CSGs, (2)
to analyze the scalability, correctness and efficiency of \tool
on large-scale real world programs.

\begin{figure}
\includegraphics[width=\linewidth]{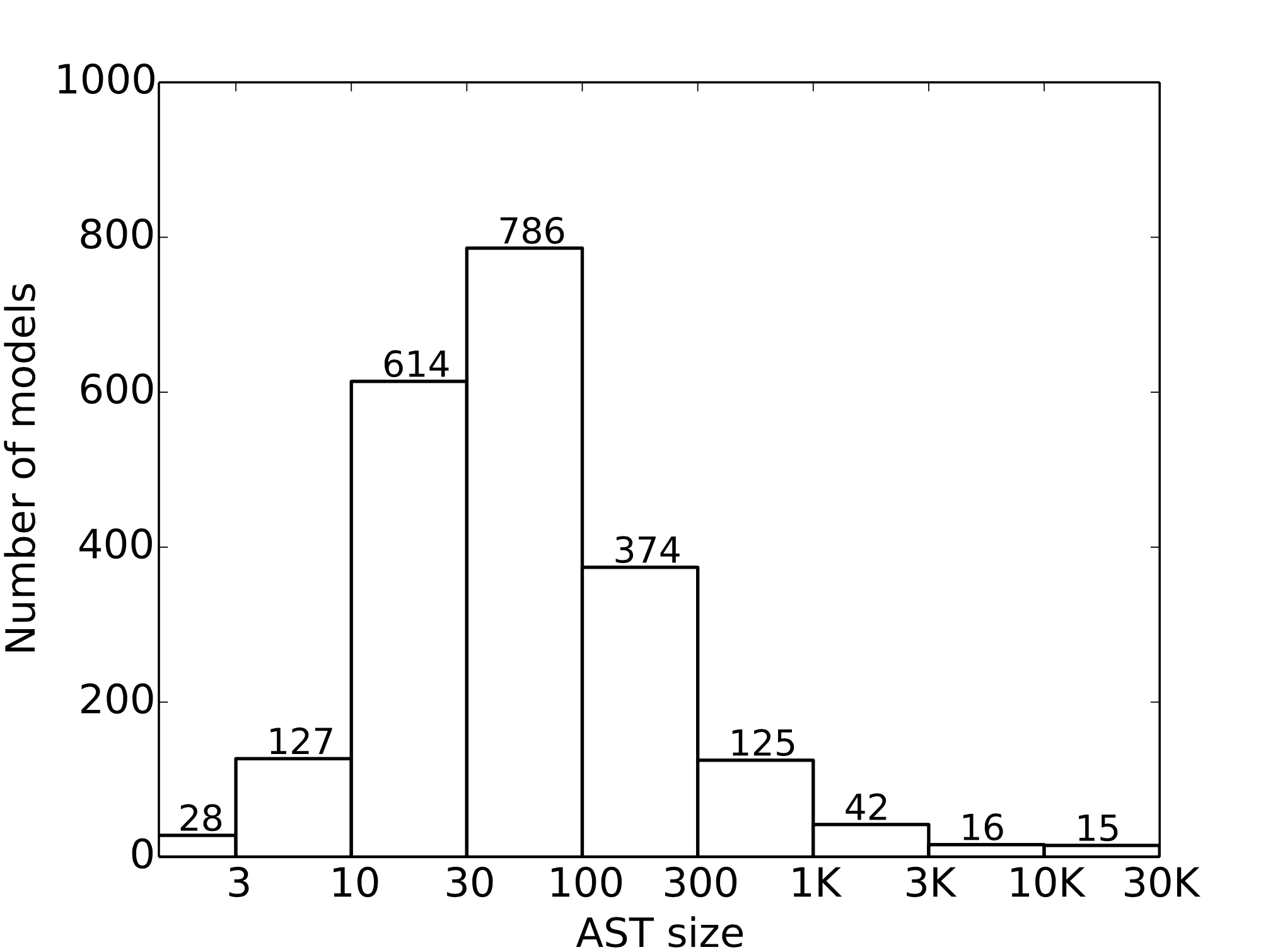}
  \caption{ \shepherd{Summary of input AST size for \tool's large scale evaluation.}}
  \label{fig:inputsize}
\end{figure}

\paragraph{Data Collection}
We built a scraper that downloaded customizable models from
Thingiverse. While most models in Thingiverse
are shared as triangular meshes which are hard to customize, models under the
"Customizable" category~\cite{customthing} are intended to be editable,
and are therefore more likely
to be accompanied with higher-level programmatic representation. Our scraper
found 12,939 OpenSCAD files from the "Customizable".
912 of these files were
invalid, i.e. they were empty, could not be compiled, or used debug features.
We filtered out files using features we do not support (like linear extrusion),
leaving 2,127 models.
Similar to \cadilac,
the OpenSCAD language supports CSG and also has features like \F{for} loops that can
be used to write more parametrizable CAD programs.
OpenSCAD can compile these programs
to flat CSG, which \tool then accepts as input.
\shepherd{Figure~\ref{fig:inputsize} summarizes the AST sizes of these inputs.}

OpenSCAD primitives like spheres and cylinders are parameterized by their
geometric precision.
The geometric precision indicates the
  quality of the mesh obtained when the CSG is compiled.
For example, a sphere with resolution 100 has a more
fine-grained mesh than a sphere with resolution 10.
We found several examples where the precision of the primitives was as high as 100.
However, OpenSCAD's compiler is slower when generating finer resolution meshes.
Since our verifier (\autoref{subsec:correct}) uses the OpenSCAD
compiler, we capped the precision of all primitives to 25.

\begin{figure*}[h]
\includegraphics[width=\linewidth]{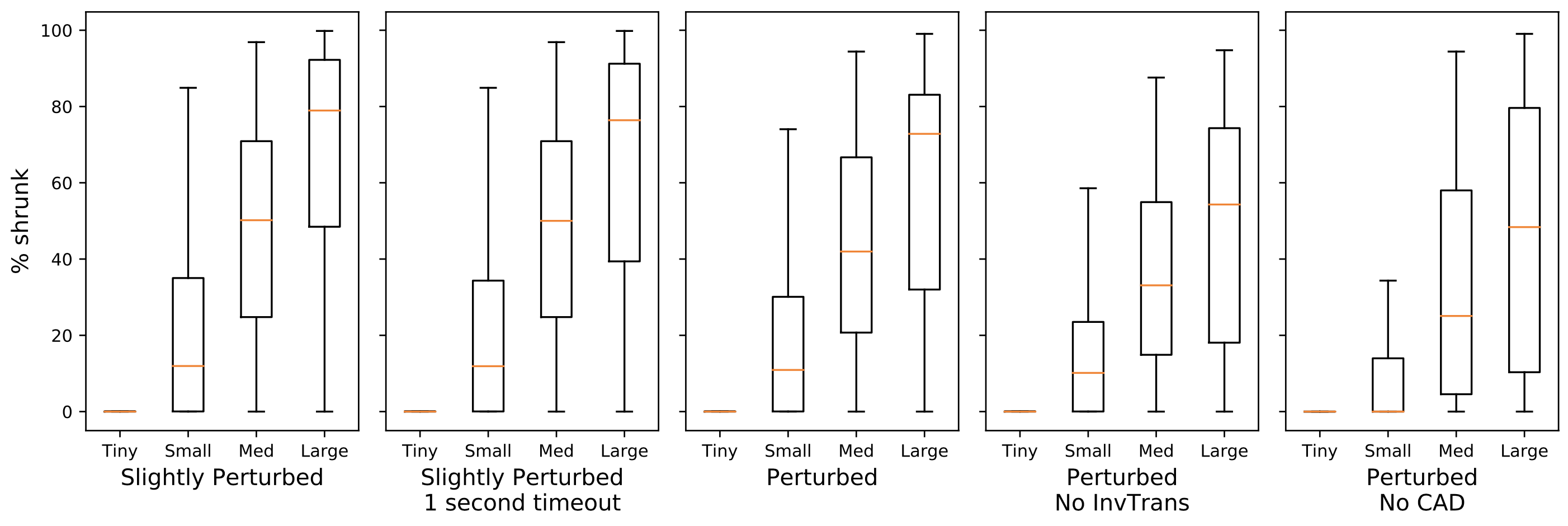}
\caption{
  Result of running \tool on 2,127 Thingiverse examples.
  Models are grouped by AST size of initial Core \caddy input:
  769 were tiny (AST size < 30),
  786 small (30 < size < 100),
  374 medium (100 < size < 300),
  and 198 large (300 < size).
}
\label{fig:thing-res}
\end{figure*}

\paragraph{Results}
\autoref{fig:thing-res} shows our results with a 60 second timeout.
We refer to the baseline result (leftmost) as slightly perturbed, as
  OpenSCAD represents affine transformations in an ambiguous way in
  its CSG format (ex: the representation of \F{Scale [-1,-1,1]} and
  \F{Rotate [0,0,180]} are identical).
The second result shows that \tool is fast; limiting it to 1 second
  has very little effect on the result.
The third result shows \tool is robust to reordering of the inputs.
The final two results show CAD rewrites or \semtag significantly
  contribute to \tool's performance.
We validated all results with the by comparing the meshes.
All Hausdorff distances were under 0.01, except for 148 cases where
  CGAL failed to compute the distance
  and we visually compared the meshes.

% then we perturbed them to reorder expressions under unions and intersections to further reflect

% \paragraph{Goals}
% Our goal in this part of the evaluation
% is to demonstrate that \tool is able to find
% \textit{parametrized}, \textit{correct}
% \cadilac programs from flat CSG programs.
% We also want to demonstrate that \tool is fast and can handle large flat programs.
% In order to analyze the quality of the parametrized programs,
% we randomly sampled \todo{XXX} programs and manually checked that \tool's result was
% indeed easier to customize than the flat CSG.
% Out of these \todo{XXX} randomly selected programs, \todo{XXX} programs were
% already originally parametrized (before we used OpenSCAD's export to CSG feature).
% We compared \tool's parametrization with the original parametrization and found that
% they compiled to equivalent meshes.
% \autoref{fig:thing-res} shows the results of our evaluation. \todo{explain}.

\subsection{Case Studies and Editability}
\label{subsec:case}

\begin{figure*}[t]
  \includegraphics[width=\textwidth]{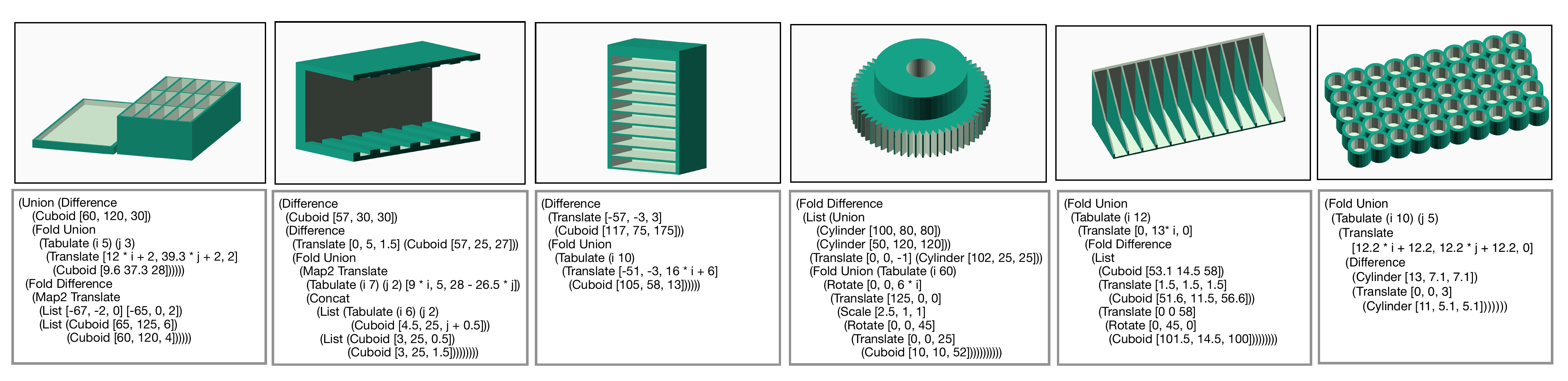}
  \caption{The first three are examples of end-to-end evaluation where \tool ran
  on the flat CSG output of a mesh decompiler~\cite{reincarnate}. The last
  three are representative examples that show the usefulness of \tool where the
  flat CSG was generated using OpenSCAD and perturbed
  to simulate  mesh decompilers.
  }
  \label{fig:cases}
\end{figure*}

This section discusses three models from the end-to-end evaluation
in \autoref{subsec:end2end}
(a fourth is illustrated in \autoref{fig:motiv})
and three models from the large scale
evaluation in \autoref{subsec:exper}.
The goal is to highlight some edits made easily possible by \tool, which
in the flat CSG (and mesh) are nearly impossible.
\autoref{fig:cases} shows a rendering of these models and the
parametrized \caddy program found by \tool.
We discuss three categories of edits.

\textit{Adding or removing components}: consider the gear shown in \autoref{fig:cases}.
Changing the tooth count in a flat CSG version of this model requires manually computing
the position of every teeth and ensuring that the spacing between them is still
equal. The \caddy program synthesized by \tool makes this modification trivial---it exposes a
function (6 $\times$ i) for \rotate
and the number of teeth ( in the \mapi), which can both be easily changed to get a different
tooth count. Adding rows or columns of components is also easy in a parametrized model. For example, in
the first model in \autoref{fig:cases}, another set of compartments can be added by changing the bounds
of \mapi.

\textit{Modifying the shape of multiple components}: in the last model in \autoref{fig:cases}, the cylinders
can be all changed to \hx by changing it in two places only. These modification in the flat CSGs require changing the shape of
each cylinder individually, which is undesirable. \autoref{fig:motiv} shows more examples of edits
where the shape of the hex-wrench holder can be changed by changing
the parameters inferred by \tool.

\textit{Applying additional affine transformations to components}:
consider the SD card rack (the second model) in \autoref{fig:cases}.
This model can be easily customized in the \caddy program to adjust the size of the slots. The \caddy program
in the figure shows that in each iteration (in \mapi), two sizes of \unit are removed from the outer box.
The dimensions of these can be changed in the function inferred for the
\unit parameters:
(\F{Cuboid [4.5, 25, j + 0.5]}) to change the slot size. Similarly, \autoref{fig:motiv} showed how
an additional rotation can be easily added to \unit to make an entirely different model.

Performing these modifications in a flat CSG is tedious and error-prone
because they require manually recomputing many parameters for multiple
components in the models. \tool makes these modifications much easier by exposing
different design parameters.

\subsection{Limitations}
\label{subsec:lim}

Some mesh decompilation tools like InverseCSG synthesize
flat CSG programs using enumerative synthesis and random sampling
based algorithms like RANSAC~\cite{inverse}. Inferring structure from
the output generated by these tools requires
equivalence under context using geometric reasoning that
our prototype currently does not support.
InverseCSG provides 50 benchmarks, on all of which we ran \tool.
The majority of the benchmarks lacked the repetitive structure \tool is
intended to infer.
For one of the models (benchmark 157, a gear),
\tool was able to infer a \polar function.
However, due to the structure of their outputs,
we had to add rewrites like:\\
\F{(Difference (Union a b) c) $\rewritesto$ (Union (Difference a c) b)}
which are unsound without a geometric solver that can check that the intersection of
\F{b} and \F{c} is empty.
We manually applied this rewrite to benchmark-157
but did not add these rewrites to
\tool's rule database due to their unsoundness.

%\todo{contextual geometric reasoning. we have shallow, context free reasnoning.}

%We also ran \tool on the flat CSGs synthesized
%by InverseCSG~\cite{inverse}.
%InverseCSG uses the RANSAC algorithm
%and enumerative search.
%\tool's results were correct,
%but it was rarely able to find any
%structure from the InverseCSG programs
%because most of InverseCSG's outputs lacked
%repetitive structure and were noisy
%due to the use of RANSAC \todo{why does it not work?
%}

%Thing-130203 is a good limitation case
%Say in the intro that none of the existing phase1 tools are good enough to be run on real things.

%     \F{Thing-64847_box_flat}    & & & & & & \\ \hline
%     \F{Thing-3148599_sdcard}  & & & & & & \\ \hline
%     \F{Thing-2921167_circle_cell}  & & & & & & \\ \hline
%     \F{Thing-3244600_cnc*} & & & & & & \\ \hline
%     \F{Thing-3907519_charger}  & & & & & & \\ \hline
%     \F{Thing-3072857_tape}& & & & & & \\ \hline
%     \F{Thing:3097951_pinheader*}& & & & & & \\ \hline
%     \F{Thing:3333935} & & & & & & \\ \hline
%     \F{Hex-wrench holder}& & & & & & \\ \hline
%     \F{} & & & & & & \\

%\input{discussion}
\section{Related Work}
\label{sec:related}

\paragraph{\egraph based Deductive Program Synthesis}

E-graphs have been used extensively in
superoptimizers~\cite{denali, eqsat, eqsat1, aiken}, and SMT solvers~\cite{z3,
lean, simplify, rosette}. \tool's core algorithm is
a generalized version of equality saturation~\cite{eqsat}.
Integrating linear solvers with compiler optimizers has a long history with
tools like Omega Calculator~\cite{pugh1, pugh2}.
Our approach of using syntactic rewrites
and an arithmetic function solver to
modify the \egraph can be considered similar to Simplify~\cite{simplify} which
uses an \egraph module for finding equivalent expressions containing
uninterpreted functions, and a Simplex module that is used for arithmetic
computations.

However, unlike \tool,
past work does not allow solvers to speculatively add potentially
profitable expressions in the \egraph.
\Semtag allows \tool to accomplish this while also
mitigating the AC-matching problem for associative and
commutative operations like list reordering and regrouping.

%The problem this
%paper tackles is that of discovering inherent structure in flat CSG-based CAD
%models.  We use an \egraph based approach to apply rewrites to a CSG that
%exposes the underlying structure. One of the contributions of our work is
%identifying that this problem has two components: an uninterpreted component
%and an arithmetic component as described in previous sections.  Our approach of
%using uninterpreted rewrites and arithmetic function solvers to modify the
%\egraph can be considered similar to Simplify~\cite{simplify} which uses an
%\egraph module for finding equivalent expressions containing uninterpreted
%functions, and a Simplex module that is used for arithmetic computations.
%\todo{mw: they do not allow solvers to speculate}

\paragraph{2D and 3D design synthesis} Nandi et al.~\cite{reincarnate} and Du
et al.~\cite{inverse} have developed tools that can decompile low-level polygon
meshes to flat CSGs.
These tools use program synthesis together with domain specific
computational geometric algorithms to discover structure in the meshes.
%\autoref{subsec:end2end} showed the results of running \tool on the outputs of
%Reincarnate~\cite{reincarnate}.
CSGNet~\cite{csgnet} uses machine learning to generate flat CSG programs for 2D
and 3D shapes.  Shape2Prog~\cite{shape} uses machine learning to infer
programs from voxel-based 3D models. They use LSTMs to infer programs with
loops. We ran \tool on the flat CSGs from both CSGNet and Shape2Prog---
since their program lengths are very small (AST depth < 7),
they are not good candidates for design parameter inference. \tool however
did find some structure in these program and generated correct
outputs.
%Our contribution, \tool is different from these tools in that it is the
%first tool that can automatically infer parameters from CSGs to generate
%high-level CAD programs in a purely functional programming language.
%Chugh et al.~\cite{chugh1}'s SKETCH-N-SKETCH is a tool
%that combines direct manipulation with programmatic manipulations for Scalable
%Vector Graphics (SVG) to detect programmatic updates based on direct
%manipulations to the SVG.
Ellis et al.~\cite{latex} developed a tool that can automatically generate
programs that correspond to hand-drawn images. They first use machine learning
to detect primitives in the drawings and then use Sketch~\cite{sketch} to find
loops and conditionals. \tool's technique is different from theirs in that they
use enumerative search to explore all programs within a given depth (their max AST depth is 3),
based on a language grammar, a specification, and a cost, whereas \tool uses a rewrite-based
synthesis technique where the specification is given as the initial CSG, and
\tool constructs an \egraph and updates it using semantics preserving rewrites.
In order to compare \tool with Ellis et al.'s~\cite{latex} tool, we
ported their 2D models to 3D and ran \tool on them. \tool's results
had similar loop structure as theirs but further comparison is not possible since their DSL is different.
Another line of work~\cite{repl} uses reinforcement
learning to synthesize programs for 2D and 3D models.
However, the programs inferred by these
approaches are much smaller compared to \tool.

In computer graphics and vision, symmetry detection~\cite{symm} in 3D shapes is
a well studied topic. It can improve performance of geometry processing
algorithms. The ability to detect folds and maps in 3D models is more general
than symmetry detection because it can find patterns in models that have
repetitive structure that is not symmetry. A simple example of this is a union of
$n$ cubes increasing in size.
In fabrication, Schulz et al.~\cite{adrianagen} developed
algorithms for optimizing parametric CAD models
using interpolation methods.
While their approach can optimize parameters,
it does not automatically infer maps and folds from flat CSG inputs.

\vspace{-0.1in}
\section{Conclusion}
\label{sec:conclusion}

%% End-user challenges in desktop manufacturing (e.g., 3D printing)
%%   have inspired recent program synthesis results
%%   that use heuristics to decompile triangle meshes to CSG.

%To help users modify low-level CAD models shared online,

This paper addresses the challenge
  of synthesizing smaller high-level CAD models
  from the noisy and unstructured outputs
  of existing triangle mesh to CSG decompilers.
%%%%  that composes with recent tools
%%%%  to decompile triangle meshes to CSG:
%%%%  given a noisy and unstructured CSG expression,
%%%%  produce an equivalent, smaller, and more editable CAD program
%%%%  with map and fold operators for expressing repetition.
We developed \emph{\tool},
  a prototype tool to synthesize \emph{\caddy} programs
  using semantics-preserving rewrites and simple solvers to ``reroll loops.''
By adapting Equality Saturation to the CAD domain,
  \tool can robustly handle common CSG variations
  exhibited by existing mesh decompilers.
\tool relies on novel \emph{\semts}
  to mitigate the AC-matching problem that
  arises when reordering CAD operations:
  solvers annotate and merge terms that
  are only equivalent modulo reordering,
  then propagate and eliminate such annotations
  through an \egraph to expose repetitive structure
  and robustly enable loop rerolling.
\Semtag are not CAD-specific;
  we are excited to explore future work
  investigating how they may be applied
  in other ordering-sensitive optimization problems,
  e.g., instruction scheduling~\cite{sched1, sched2}.

To the best of our knowledge,
  \tool is the first tool of its kind.
%%We qualitatively evaluated \tool in case studies:
%%  to highlight design strengths and
%%  some weaknesses which require contextual rewrites
%%  \tool does not yet support.
We performed an early survey of
  \thingBenchCount real-world CAD models from Thingiverse.
Our evaluation shows that \tool can
  dramatically shrink many CAD models in seconds.

In future work,
  we are excited to explore richer rewrites for
  contextual equivalence (\autoref{subsec:lim}),
  more expressive cost functions for capturing
  richer notions of editability,
  and connections to interactive CAD editing
  using direct manipulation tools like Sketch-n-Sketch~\cite{chugh1}.

\bibliography{reference}

%% Appendix
%\appendix
%\input{appendix}

%Text of appendix \ldots

\end{document}